 \documentclass[journal,10pt,onecolumn,draftclsnofoot,]{IEEEtran}
\usepackage{graphicx,cite}
\usepackage{epstopdf}

\usepackage{amssymb,amsmath,bm}
\usepackage{amsmath}
\usepackage{textcomp}
\usepackage{algorithm}
\usepackage{algorithmic}
\usepackage{epsfig}
\usepackage{slashbox}
\usepackage[lofdepth,lotdepth]{subfig}
 \usepackage{multirow}
\usepackage{enumerate}
\usepackage{mathrsfs}
\usepackage{comment}
\newtheorem{thm}{Theorem}
\newtheorem{prop}[thm]{Proposition}
\newtheorem{lem}[thm]{Lemma}
\newtheorem{remark}{Remark}

\newtheorem{definition}{Definition}
\newtheorem{cor}[thm]{Corollary}
\newtheorem{note}{Note}
\newcommand*{\QEDB}{\hfill\ensuremath{\square}}
\newcommand{\RN}[1]{%
  \textup{\uppercase\expandafter{\romannumeral#1}}%
}

\title{$H_{\infty}$ Optimal Control of Jump Systems Over Multiple Lossy Communication Channels}
\author{\IEEEauthorblockN{Abhijit Mazumdar, Srinivasan Krishnaswamy and Somanath Majhi}\\
\IEEEauthorblockA{Dept. of Electronics and Electrical Engineering \\
Indian Institute of Technology Guwahati, Guwahati-781039, India \\
{ \tt \{abhijit.mazumdar, srinikris, smajhi\}@iitg.ac.in}
}}

\author{Abhijit Mazumdar, Srinivasan Krishnaswamy and Somanath Majhi% <-this % stops a space
	\thanks{Abhijit Mazumdar, Srinivasan Krishnaswamy and Somanath Majhi are with the Department of Electronics \& Electrical Engineering,
		Indian Institute of Technology Guwahati, Guwahati-781039, India. 
	   \small E-mails: \{abhijit.mazumdar, srinikris, smajhi\}@iitg.ac.in}
}
\begin{document}
\vspace{-2cm}
  \maketitle
  \begin{abstract}
 In this paper, we consider the $H_{\infty}$ optimal control problem for a Markovian jump linear system (MJLS) over a lossy communication network. It is assumed that the controller communicates with each actuator through a different communication channel. We solve the $H_{\infty}$ optimization problem for a Transmission Control Protocol (TCP) using the theory of dynamic games and obtain a state-feedback controller. The infinite horizon $H_{\infty}$ optimization problem is analyzed as a limiting case of the finite horizon optimization problem. Then, we obtain the corresponding state-feedback controller, and show that it stabilizes the closed-loop system in the face of random packet dropouts.
\end{abstract}
\vspace{-0.5cm}
\section{Introduction}
\label{intro} 
\par A network control system (NCS) is a system that uses a communication network to share information among different subsystems, viz: controller, actuators, and sensors.  NCS has many practical applications  
 \cite{oncu2012string,millan2014formation,yao2015wide}.
\par A communication network introduces time-delays \cite{f}, packet losses \cite{g}, and quantization \cite{h} into the system. Packet loss is a very serious issue as it can bring about system instability. As Bernoulli processes are easy to analyze, packet loss is often modeled as an independent and identically distributed (i.i.d.) Bernoulli process \cite{g}, \cite{22}. However, this doesn't capture the temporal correlation often seen in packet losses. One can alternatively use the Gilbert-Elliot channel model, wherein packet losses are modeled as a Markov process \cite{j}.  
\par There are two types of communication protocols used in NCSs: Transmission Control Protocol (TCP) and User Datagram Protocol (UDP). In a TCP-like protocol, packet receptions are acknowledged, while in a UDP-like protocol there is no such acknowledgment. 
\par In many practical systems, for example dc-dc converters, random factors such as environmental changes and component failure can bring about abrupt changes in the system behavior \cite{vargas2016optimal}. To factor this in, one can consider a set of mathematical models. Each abrupt change causes the system behavior to switch from one model to another. A system where these mathematical models are linear and the switching process is a Markov process is known as a Markovian jump linear system \cite{costa2006discrete}.
\par
 Linear quadratic Gaussian (LQG) control problem for an NCS with a TCP-like protocol is dealt with in \cite{g,bruno}. In \cite{g}, an LQG controller with a UDP-like protocol is also designed. Extending this work, \cite{j} studies the LQG control problem with Markovian packet losses. The LQG control over multiple lossy networks under a TCP-like protocol is investigated in \cite{garone}. In \cite{abhijit_mtns}, a linear quadratic optimal controller for an MJLS is designed over multiple Gilbert-Elliott type channels.
\par
The $H_{\infty}$ control problem with a Bernoulli packet loss model has been addressed in \cite{p,wang2007robust,18}. In \cite{wang2013h}, a time-invariant $H_{\infty}$ controller is designed with a Markovian packet loss model. With a TCP-like protocol, a minimax ($H_{\infty}$ optimal) control problem  with a Bernoulli packet loss model is investigated in \cite{22,i}. For both TCP-like and UDP-like protocols, minimax controllers are designed in \cite{m}. \cite{moon2014minimax} generalizes the results of \cite{i} to the multi-channel case. The $H_{\infty}$  optimal control problem for a linear time-invariant (LTI) system over a Gilbert-Elliott channel is solved in \cite{mazumdar2017h}.
 %Control of jump systems over communication networks has been investigated in \cite{liu2009stabilization,zhang2013robust,dong2013distributed,zhang2017distributed,abhijit_mtns,shen2018quantized}.
 In \cite{liu2018dynamic}, an $H_{\infty}$ controller for an MJLS with Bernoulli type packet losses in the measurement channel is designed.
\par In this paper, we solve the optimal $H_{\infty}$ control problem for an MJLS over multiple lossy channels. The feedback channels are assumed to be lossless \cite{kawan2016network}. In the event of a packet loss, we employ the zero input strategy.  It is assumed that each actuator communicates with the controller through a unique Gilbert-Elliott type channel.
 Existence conditions for the finite horizon controller are derived in terms of the disturbance attenuation level $\gamma$ and the packet arrival probabilities. The convergence of the infinite horizon cost function along with the stability analysis are also investigated. Using a numerical example, we demonstrate the convergence of the infinite horizon cost and its dependence on the control packet arrival probabilities. 
 \par
 The paper is structured as follows. Section II describes the $H_{\infty}$ control problem with packet losses. Section III contains the solution to the problem for both finite and infinite horizon cases. The convergence of the infinite horizon cost function and the stability of the closed-loop system are also investigated. Simulation results are presented in Section IV followed by the conclusion in section V.  
    \\
     \textbf{\textit{Notation:}} $Pr(.)$ is used to denote probability and $Pr(.| T)$ represents conditional probability given T. $\textbf{diag}\{a_1,...,a_n\}$ represents a diagonal matrix with $a_1,...,a_n$ as its diagonal elements. $\mathbb{E}[.]$ stands for expected value of a random variable, while $\mathbb{E}[.|Y]$ denotes expected value given $Y$. $||x||:=(x^Tx)^{1/2}$ and $||x||_P:=(x^TPx)^{1/2}$ denote the Euclidean norm and the weighted Euclidean norm respectively. For matrices $X(0), X(1),...,X(l)$, $\Pi_{c=0}^l X(c):=X(l)X(l-1)...X(0)$.
      $\mathcal{L}_2([0,\infty),\mathbb{R}^n)$ is the space of square-summable sequences of vectors $\{v_k\}$. $x_{l:l+n}$ denotes the sequence $\{x_l,x_{l+1},...,x_{l+n}\}$. $v^{1:n}$ represents the sequence $\{v^1,v^2,...,v^n\}$. For a matrix $P$, $P>0$ and $P\geq 0$ imply that $P$ is positive definite and positive semi-definite respectively. Further, by $P<\infty$, we imply that all the elements of $P$ are finite.    %${u}^{0:k}=\{u_0,u_1,...,u_k\}$ and ${w}^{0:k}=\{w_0,w_1,...,w_k\}$.                 
   \section{Problem Formulation} 
    %Let $(\Omega, \mathcal{F}, P)$ be the given probability space. 
    Consider the following discrete-time Markovian jumped linear system:
    \small
    \begin{equation}
    \begin{split}
    & x_{k+1}=A(r_k)x_k+B(r_k)u_k^a+D_1(r_k)w_k\\
    & z_k=C(r_k)x_k+D(r_k)u_k^a,
    \end{split}
    \label{system_eq}
    \end{equation}
    \normalsize
     where $x_k$ ${\in}$ $\mathcal{X}$ $\subseteq$ $\mathbb{R}^n$ denotes the state vector, $u_k^a$ ${\in}$ $\mathcal{U}$ $\subseteq$ $\mathbb{R}^m$ denotes the control input to the actuators, $w_k$ ${\in}$ $\mathcal{W}$ $\subseteq$ $\mathcal{L}_2([0,{\infty}),\mathbb{R}^s)$ denotes the disturbance input, $z_k$ ${\in}$ $\mathbb{R}^p$ is the controlled output. $\{r_k\}$ is an irreducible, aperiodic and time-homogeneous Markov chain where $r_k$ $\in$ $\mathcal{D}\triangleq \{1,2,...,\mathcal{M}\}$ ($\mathcal{M}<\infty$). The transition probability matrix of the Markov chain is given by $\mathcal{T}=[p_{ij}]$, where 
    \small
    \begin{equation*}
    p_{ij}=Pr(r_{k+1}=j|r_k=i); \forall i,j \in \mathcal{D}, k=0,1,2,...
    \end{equation*}
    \normalsize
   \par Standard terminology related to a Markov chain is used in this work.
 % \begin{enumerate}[(i)]
  %\item  A Markov chain is said to be \textit{time-homogeneous} if $P(r_{k+n}=j \Big| r_{n}=i)=P(r_{l+r}=j \Big| r_{r}=i)$; $\forall k,n,r,l \in \mathbb{Z}^+$.
 % \item A state $j$ is \textit{accessible} from state $i$ (written as $i \rightarrow j$) if there exists an integer $n({ij})\geq 0$ such that
 % \small \begin{equation*}
  % Pr(r_{n({ij})}=j|r_0=i)>0. %\hspace{0.2cm} \textrm{for} \hspace{0.2cm} \textrm{some} \hspace{0.1cm} \textrm{finite} \hspace{0.1cm} k\in \mathbb{Z}^+.
  %\end{equation*} 
  %\normalsize
  %\item A state $i$ is said to \textit{communicate} with state $j$ if both $i \rightarrow j$ and $j \rightarrow i$.
  %\item A \textit{communicating class} is a maximal set of states where every state communicates with each other.
  %\item A communicating class is \textit{closed} if there does not exist any transition with nonzero probability from the class to a state outside it. 
   %\item A Markov chain is \textit{irreducible} if its state space is a single communicating class.
  % \item A state $i$ is said to be a \textit{aperiodic} or has period $1$ if $p_{ii}>0$.
  % \item A Markov chain is \textit{aperiodic} if all of its states are aperiodic. 
  % \item Suppose $r_0=i\in \mathcal{D}$. Consider a random variable $\mathcal{H}_i$ such that $\mathcal{H}_i=\textrm{inf}\Big{\{}n\geq 1: r_n=i \Big{\}}$. Now, the state $i$ is called \textit{recurrent} if 
  % \begin{equation*}
   %	Pr\Big(\mathcal{H}_i<\infty \Big| r_0=i\Big)=1.
  % \end{equation*}
  % \item A Markov chain is called \textit{recurrent} if all of its states are recurrent.
  %\end{enumerate}
    \par Throughout the paper, it is assumed that state of the system $x_k$ and Markov chain state $r_k$ are directly accessible to the controller. We also assume that $A(r_k)$ is full rank for all $r_k \in \mathcal{D}$. 
    \\  Suppose $u_k$ is the controller output which is sent to the actuators through the lossy network. If $u_k^a$ is the input to the actuators, then:
    \small
\begin{equation}
  u_k^a=\xi_ku_k,
  \label{multi_channel_u}
\end{equation}    
\normalsize
    where  $\xi_k=\textbf{diag}\{v^1_{k},v^2_{k},...,v^m_{k}\}$ represents the packet loss conditions of all the channels at the time-index $k$. For each $i \in \{1,2,..,m\}$, $v^i_{k}$ denotes the packet loss condition in the $i^{th}$ channel ($v_k^i=0$ implies a packet loss and $v_k^i=1$ implies a successful packet delivery in the $i^{th}$ channel). The Gilbert-Elliott channel model is a two-state Markov chain, successful delivery of a packet being the good state and packet loss being the bad one \cite{j} \cite{l}. Let $\bar{v}^l \hspace{0.1cm}=\hspace{0.1cm} Pr(v^l_{k}=1\big| v^l_{k-1}=1)$, $\bar{\mu}^l \hspace{0.1cm}=\hspace{0.1cm} Pr(v^l_{k}=1\big| v^l_{k-1}=0)$.\\
   % The transition probability matrix of the model is given by: 
%     For the multi-channel case, \textit{Note 1} transform to the following: 
\begin{note} \label{note_1}
	At $k=0$, the probability of a data packet loss (or, successful packet arrival) in the $l^{th}$ channel is given as follows  \cite{j}: 
	\small
	\begin{equation*}
	\begin{split}
	Pr\big(v^l_{k}=0\big)=(1-\bar{v}^l)/(1+\bar{\mu}^l -\bar{v}^l) \\  Pr\big(v^l_{k}=1\big)= \bar{\mu}^l/(1+\bar{\mu}^l -\bar{v}^l).
	\end{split}
	\end{equation*} \QEDB
	\normalsize
\end{note}
  \begin{note} \label{note_2}
  	It is to be noted that $\{r_k\}$ and $\{v_k^l\}$ for each $l\in\{0,1,...,(2^m-1)\}$ are independent Markov processes. \QEDB 
  	\par For a TCP-like protocol, the information set $\mathcal{I}_k$ available to the controller at $k^{th}$ time-index is expressed as: 
  	\small
  	\begin{equation*}
  	\begin{split}
  	& \mathcal{I}_k=\{x_0,...,x_k,r_0,...,r_k,\xi_0,...\xi_{k-1}\}.
  	\end{split}
  	\end{equation*}
  	\normalsize
  \end{note}
\par The control policy $\zeta_{0:k}$ and the disturbance policy $\eta_{0:k}$ for a horizon $k$ are the sequences $\zeta_{0:k}=\{\zeta_0,...,\zeta_k\}$ and $\eta_{0:k}=\{\eta_0,...,\eta_k\}$, respectively. Here, $\zeta_i$ maps the information set $\mathcal{I}_i$ to the control input at the $i^{th}$ time-index, i.e., $u_i=\zeta_i(\mathcal{I}_i)$. Similarly, $\eta_i$ maps the information set $\mathcal{I}_i$ to the disturbance input at the $i^{th}$ time-index, i.e., $w_i=\eta_i(\mathcal{I}_i)$.  $\zeta_k^*$ and $\eta_k^*$ are the optimal control and disturbance policy, respectively.
\\In this work, we consider the following notion of stability.
 \begin{definition} The system (\ref{system_eq}) with $u_k\equiv 0$, $w_k\equiv 0$ is said to be mean-square stable if  $\underset {k \rightarrow \infty} {\textrm{lim}} \mathbb{E} \Big[ ||x_k||^2 \Big| \mathcal{I}_0 \Big] = 0$, for all $x_0$ and $r_0$. \QEDB
\end{definition}
 In \cite{ji1988controllability}, the following notion of observability for an MJLS is defined.
\begin{definition}  Consider the system (\ref{system_eq}) without disturbance ($w_k\equiv 0$).  Take any initial Markov process state $r_0$, and any two initial system states $x_0^1$ and $x_0^2$. Suppose, for a known input $u_k$, $z_k(x_0=x_0^1)=z_k(x_0=x_0^2)$, $\forall k\leq T$ implies   $Pr(x_0^1=x_0^2)>0$. The system is said to be weakly observable if $\mathbb{E}\Big[T\Big]<\infty$. \QEDB
\end{definition} 
Further, to test weak observability,  \cite{ji1988controllability} also provides an algebraic condition as given by lemma below.
\begin{lem} \label{lem_oo} The system (\ref{system_eq}) without disturbance ($w_k \equiv 0$) is said to be weakly observable if and only if there exists a transition path $\{r_0, r_1, ..., r_{T-1}\}$ inside $\mathcal{D}$ with $T<\infty$, for which the jump observability matrix $\mathcal{O}({r_0,r_1,...,r_{T-1}})$ has 
	\small
	\begin{equation*}
	\textrm{rank} \hspace{0.1cm} \mathcal{O}({r_0,r_1,...,r_{T-1}})= \hspace{0.1cm}
	\textrm{rank} \hspace{0.1cm} \begin{aligned}
	\begin{bmatrix}
	C(r_0)  \\
	C(r_1)A(r_0)\\
	.\\
	.\\
	.\\
	C(r_{T-1}){\Pi}_{i=0}^{T-2}A(r_i)  
	\end{bmatrix}=n.
	\end{aligned}
	\end{equation*} \QEDB
	\normalsize
\end{lem}  
\begin{remark}  In \cite{ji1988controllability}, the condition for weak observability is presented for the case when the Markov chain has more than one closed communicating class. In our case, as the Markov chain $\{r_k\}$ is irreducible,  we have only one closed communicating class $\mathcal{D}$. Hence, the system will be weakly observable if there exists a transition path $\{r_0, r_1, ..., r_{T-1}\}$ inside $\mathcal{D}$ with $T<\infty$ for which the condition given in Lemma \ref{lem_oo} is satisfied. \QEDB
\end{remark}
%\begin{remark}  In \cite{ji1988controllability}, the condition for weak observability is presented for the case when the Markov chain has more than one closed communicating class. In our case, as the Markov chain $\{r_k\}$ is irreducible,  we have only one closed communicating class $\mathcal{D}$. Hence, the system will be weakly observable if there exists an initial Markov state $r_0 \in \mathcal{D}$ for which the condition given in Lemma \ref{lem_oo} is satisfied. 
%\end{remark}
 \par The objective of this work is to solve the following problem.
\par \textit{Problem:} Design state-feedback control policies for the system (\ref{system_eq}) with network induced constraint (\ref{multi_channel_u}) such that with a state-feedback control law, the closed-loop system attains the following requirements:
\begin{enumerate}[(R.1)]
	\item  $\mathcal{L}_2$ gain from the disturbance input $w_k$ to the controlled output $z_k$ must be less than or equal to some $\gamma>0$, i.e., with zero initial condition $x_0=0$,
	\small
	\begin{equation*}
	\sum_{k=0}^{N} \mathbb{E} \Big[ ||z_k||^2 \Big| \mathcal{I}_0 \Big] \leq  \gamma^2 \sum_{k=0}^{N}||w_k||^2, \forall N \in \mathbb{Z}^+.
	\end{equation*}
	\normalsize
	\item The closed loop system is mean-square stable. 
\end{enumerate}
 \par The $H_{\infty}$ optimal control problem can be analyzed in the framework of dynamic games \cite{m,n}. Subject to the constraints defined by the system dynamics (\ref{system_eq}), one can formulate a zero-sum game with the following cost function:
 \small
 \begin{equation}
J_N({\zeta}_{0:N-1},{\eta}_{0:N-1})=\mathbb{E} \Big[||x_{N}||_{W}^2+\sum_{k=0}^{N-1} ||z_k||^2-\gamma^2||w_k||^2 \Big| \mathcal{I}_0 \Big].
\label{cost_fun1}
\end{equation} 
\normalsize
Using Equation (\ref{system_eq}) in (\ref{cost_fun1}), the cost function becomes:
\small
\begin{equation}
 \begin{split}
J_N({\zeta}_{0:N-1},{\eta}_{0:N-1})=\mathbb{E} \Big[||x_{N}||_{W}^2+\sum_{k=0}^{N-1} ||x_k||_{W({r_k})}^2\\+||u_k^a||_{R({r_k})}^2-\gamma^2||w_k||^2\Big| \mathcal{I}_0\Big] ,
\end{split}
\label{cost_fun2}
\end{equation} 
\normalsize
\\where $W\geq 0$, $W(r_k)=C^T(r_k)C(r_k)$ and $R(r_k)=D^T({r_k})D({r_k})$. Further, $C(r_k)$ and $D(r_k)$ satisfy the following assumptions:
\begin{enumerate}[(a)]
 \item $C^T({r_k})D({r_k})=0, \forall r_k \in \mathcal{D}$; implying that there are no cross product terms in the cost function (\ref{cost_fun2}).
 \item $R(r_k)>0$, $ \forall r_k \in \mathcal{D}$; implies nonsingularity of the  
optimal control problem.
\end{enumerate} 
\par In the game with the cost function (\ref{cost_fun2}), the control input ${u}_k$ acts as the minimizing player and the disturbance ${w}_k$ acts as the maximizing player. From the theory of zero-sum game, it is well established that the game admits a minimax solution or Nash equilibrium, if one can find a saddle-point policy $({\zeta}^*_{0:N-1},{\eta}^*_{0:N-1})$ which satisfies the inequality:
\small
\begin{equation}
J_N({\zeta}^*_{0:N-1},{\eta}_{0:N-1}) \leq J_N({\zeta}^*_{0:N-1},{\eta}^*_{0:N-1}) \leq J_N({\zeta}_{0:N-1},{\eta}^*_{0:N-1}).
\end{equation}
\normalsize
  %We shall observe that the saddle point conditions will depend explicitly on control packet arrival probabilities and $\gamma$.  
 \begin{remark}
  For the particular case when $\mathcal{D}=\{1\}$, i.e., for an LTI system, this problem has been addressed in \cite{mazumdar2017h}. However, only one communication channel has been considered between the actuators and the control unit. \QEDB
 \end{remark} 
\begin{remark} If $\xi_k$ takes only two values: $0_{m\times m}$ or $1_{m\times m}$, then multi-channel case becomes equivalent to the single channel case. \QEDB
\end{remark}
\par Before going on to the main results, we define a few terms which will be used in the sequel.
\begin{enumerate}[(a)]
	\item $\mathscr{G}=\{1,2,...,m\}$ is the index set for the actuators.
	\item $\mathscr{I}_j$ is given by:
	\begin{equation*}
	\begin{split}
	\mathscr{I}_j=\Big{\{} p\in \mathscr{G}: \hspace{0.2cm} &   p^{th} \hspace{0.1cm} \textrm{entry in the $m$-bit binary  } \\
	& \textrm{ representation of $j$ is $1$,  where }  \\
	& \textrm{ position   number of the least}\\
	& \textrm{significant bit (LSB)  is assumed to }\\
	& \textrm{ be 1} \Big{\}}.
	\end{split}
	\end{equation*} 
	\normalsize
	%\item $\bar{v}_l=P(v_{k,l}=1|v_{k-1,l}=1)$, $\bar{\mu}_l=P(v_{k,l=1}|v_{k-1,l}=0)$; for all $l\in \{1,2,...,m\}$. 
	\item $
	\mathcal{N}(j)=\textbf{diag}\big{\{}a_{ll} \big{\}}; \hspace{0.05cm}  \textrm{where} \hspace{0.2cm} 
	a_{ll}=\begin{cases} 1, & \mbox{if} \hspace{0.2cm} l \in \mathscr{I}_j \\
	0, & \mbox{if}   \hspace{0.2cm} l \notin \mathscr{I}_j
	\end{cases}\\
	\textrm{for} \hspace{0.2cm} l=1,2,...,m \hspace{0.05cm}
	$.
	\item Suppose, at the stage $(k-1)$ the actuators that receive control command are the ones indexed by the elements of $\mathscr{I}_j$. Then, the probability that, at the stage $k$, the actuators that receive control command are those indexed by the elements of $\mathscr{I}_l$ is given by:
	\\ $\mathcal{P}^j(l)=\underset{h\in \mathscr{I}_l}{\Pi}Pr\big(v^h_{k}=1\big| \xi_{k-1}=\mathcal{N}(j)\big)\underset{h\notin \mathscr{I}_l}{\Pi}Pr\big(v^h_{k}=0\big| \xi_{k-1}=\mathcal{N}(j)\big)$.\\
	For the case when there is no information available regarding the previous packet loss condition (e.g. at the stage $k=0$), we define the following function:
	\small
	\begin{equation*}
	\hat{P}(l)= \underset{h\in \mathscr{I}_l}{\Pi}Pr(v^h_{k}=1)\underset{h\notin \mathscr{I}_l}{\Pi}Pr(v^h_{k}=0).
	\end{equation*}
	\normalsize
	Also, $\mathcal{P}_k(l)=\mathcal{P}^j(l)$ for $k\geq 1$ and $\xi_{k-1}=\mathcal{N}(j)$.
	%\small
	%\begin{equation*}
	%\mathcal{P}_k(l)=\mathcal{P}^j(l).	
	%\end{equation*}
	%\normalsize
	 Further, $\mathcal{P}_0(l)=\hat{\mathcal{P}}(l)$.
	\item Let ${\mathbf{Y}}(.)$ be a map from $2^{\mathscr{G}}$ to spaces which are closed under addition and scalar multiplication.
	Now, $\mathbf{L}^j(.)$, $\hat{\mathbf{L}}(.)$ and $\mathbf{L}_k(.)$ are defined as follows:
	\small
	\begin{equation*}
	{\mathbf{L}}^j\Big({\mathbf{Y}}(.)\Big) =\sum_{l=0}^{2^{m}-1} \mathcal{P}^j(l)\mathbf{Y}(l) 
	\label{L_i}.
	\end{equation*}
	\small
	\begin{equation*}
	{\hat{\mathbf{L}}}\Big({\mathbf{Y}}(.)\Big) =\sum_{l=0}^{2^{m}-1} \hat{\mathcal{P}}(l)\mathbf{Y}(l) 
	\label{L_i_0}.
	\end{equation*}
	For $k\geq 1$, $\mathbf{L}_k(.)=\mathbf{L}^j\Big({\mathbf{Y}}(.)\Big)$ if $\xi_{k-1}=\mathcal{N}(j)$, and $\mathbf{L}_0(.)=\hat{\mathbf{L}}\Big({\mathbf{Y}}(.)\Big)$.
\end{enumerate} 
\normalsize
 %In the multi-channel case, partial information loss is possible which is not the case with the single channel case. 
\vspace{-0.2cm}
 \section{Main Results}   
%\vspace{-0.2cm}
In this section, we deal with the design of finite horizon and infinite horizon controllers.
%\vspace{-9cm}
%\vspace{-10cm}
\vspace{-0.5cm}
 \subsection{Finite Horizon Control:}
By substituting (\ref{multi_channel_u}) in cost function (\ref{cost_fun2}), we get the following:
\small
\begin{equation}
 \begin{split}
J_N(\zeta_{0:N-1},\eta_{0:N-1})=\mathbb{E} \Big[x_N^TWx_N+\sum_{k=0}^{N-1} x_k^TW(r_k)x_k\\+u_k^T \xi_k^TR(r_k)\xi_ku_k-\gamma^2w_k^Tw_k \Big| \mathcal{I}_0 \Big] .
\end{split}
\label{cost_fun_multi_channel}
\end{equation}
\normalsize
 The optimal cost-to-go or value function at the stage $k$ is given by:
 \small
 \begin{equation} 
 \begin{split}
 &V_{k,N}(x_k,r_k,\xi_{k-1})= \underset {u_{k:N-1}}{\textrm{min}} \hspace{0.1cm} \underset {w_{k:N-1}} {\textrm{max}} \hspace{0.1cm}\mathbb{E} \hspace{0.1cm} \Big[ x_N^TWx_N\\
 &+\sum_{p=k}^{N-1} x_p^TW(r_p)x_p+u_p^T \xi_p^TR(r_p)\xi_pu_p-\gamma^2w_p^Tw_p \Big| \mathcal{I}_k\Big].
 \end{split}
 \label{value_fn1_multi_1}
 \end{equation}
 \normalsize
 Using principle of optimality, one can express (\ref{value_fn1_multi_1}) as:
 \small
 \begin{equation} 
 \begin{split}
 V_{k,N}(x_k,r_k,\xi_{k-1}) = \underset {u_k}{\textrm{min}} \hspace{0.1cm} \underset {w_k} {\textrm{max}} \hspace{0.1cm}\mathbb{E} \hspace{0.1cm} \Big[ x_k^TW(r_k)x_k+u_k^T \xi_k^TR(r_k)\xi_ku_k\\-\gamma^2w_k^Tw_k +V_{k+1,N}(x_{k+1},r_{k+1},\xi_k)|\mathcal{I}_k\Big].
 \end{split}
 \label{value_fn1_multi}
 \end{equation}
 \normalsize  
Equation (\ref{value_fn1_multi}) is called the Isaacs equation (finite horizon).\\
We now proceed to derive conditions under which the value of the game with cost function (\ref{value_fn1_multi}) has a well defined solution.
\par For $k\in[0,N-1]$, $j \in \{0,1,...,(2^m-1)\}$, $r_k=i\in \mathcal{D}$, consider the following coupled algebraic Riccati equations (CAREs):
\small
\begin{equation}
	\begin{split}
		&\Xi_{k,N}(i,j)= W(i)+ \Gamma_{k,N}^T(i,j)\mathbf{L}_k\Big(\mathscr{Q}(i,.)\Big)\Gamma_{k,N}(i,j)\\
		& -\gamma^2\Psi_{k,N}^T(i,j)\Psi_{k,N}(i,j)\\
		&  +\sum_{l=0}^{2^m-1}\mathcal{P}_k(l)\mathbb{E}\Big[\Big(A(i)-B(i)\mathcal{N}(l)\Gamma_{k,N}(i,j)+D_1(i)\Psi_{k,N}(i,j)\Big)^T\\
		& \times \mathscr{X}_{k+1,N}(i,l) \Big(A(i)-B(i)\mathcal{N}(l)\Gamma_{k,N}(i,j)+D_1(i)\Psi_{k,N}(i,j)\Big)\Big],
	\end{split}
	\label{care_multi}
\end{equation}
\normalsize
where, 
\begin{subequations}
	\small
	\begin{equation}
		\begin{split}
			& \Gamma_{k,N}(i,j)\\
			& =\Big(\Lambda_{k,N}(i,j)\Big)^{-1}\Big[\mathscr{\mathbf{L}}_k\Big(\mathscr{T}_{k+1,N}(i,.) \Big) \Big(A(i)+D_1(i)\Psi_{k,N}(i,j) \Big)\Big],
			\label{Gamma_multi}
		\end{split}
	\end{equation}
	\normalsize
	\small
	\begin{equation}
		\begin{split}
			& \Psi_{k,N}(i,j)\\
			& =\Big{[}I+\big(\Theta_{k,N}(i,j)\big)^{-1}D_1^T(i)\mathscr{\mathbf{L}}_k\Big(\mathscr{T}^T_{k+1,N}(i,.)\Big)\\
			& \times \big(\Lambda_{k,N}(i,j)\big)^{-1} \mathscr{\mathbf{L}}_k\Big(\mathscr{T}_{k+1,N}(i,.)\Big)D_1(i) \Big{]}^{-1} \Theta_{k,N}^{-1} (i,j)\\
			& \times \Big{[}D_1^T(i)\mathscr{\mathbf{L}}_k\Big(\mathscr{X}_{k+1,N}(i,.)\Big) -D_1^T(i)\mathscr{\mathbf{L}}_k\Big(\mathscr{T}^T_{k+1,N}(i,.)\Big)\\
			& \times \Big(\Lambda_{k,N}(i,j)\Big)^{-1} \mathscr{\mathbf{L}}_k\Big(\mathscr{T}_{k+1,N}(i,.)\Big) \Big{]}A(i),
		\end{split}
		\label{Psi_multi}
	\end{equation}
	\begin{equation}
		\begin{split}
			& \Theta_{k,N}(i,j) =\gamma^2I-D_1^T(i)\mathscr{\mathbf{L}}_k\Big(\mathscr{X}_{k+1,N}(i,.)\Big)D_1(i),
		\end{split}
		\label{theta}
	\end{equation}
	\begin{equation}
		\begin{split}
			& \Lambda_{k,N}(i,j) =\mathbf{L}_k\Big(\mathscr{R}_{k+1,N}(i,.)\Big),
		\end{split}
	\end{equation}
	\begin{equation}
		\mathscr{Q}(i,e)=\mathcal{N}(e) R(i) \mathcal{N}(e),
	\end{equation}
	\begin{equation}
		\mathscr{T}_{k+1,N}(i,e)=\mathcal{N}(e) B^T(i)\mathscr{X}_{k+1,N}(i,e),
	\end{equation}
	\begin{equation}
		\mathscr{R}_{k+1,N}(i,e)=\mathcal{N}(e) \big[R(i)+B^T(i)\mathscr{X}_{k+1,N}(i,e)B(i)\big] \mathcal{N}(e),
	\end{equation}
	\begin{equation}
	\mathscr{X}_{k+1,N}(i,l)=\sum_{d=1}^{\mathcal{M}} p_{id} \Xi_{k+1,N}(d,l),
	\end{equation}
	\begin{equation}
	\begin{split}
		\textrm{with} \hspace{0.2cm} \Xi_{N,N}(i,j)=W; \hspace{0.2cm} \forall i\in \mathcal{D}, j\in \{0,1,...,(2^m-1)\}.
		\end{split}
	\end{equation}
\end{subequations}
\normalsize
\begin{note} \label{note_aa}
It is assumed that $\Xi_{q,N}(i,j)=0$ for all $q\geq N+1$, $i\in \mathcal{D}$, $j\in \{0,1,...,(2^m-1)\}$. Thus for all $i\in \mathcal{D}$, $j\in \{0,1,...,(2^m-1)\}$, $s\geq N$, one gets $\Theta_{s,N}(i,j)=\gamma^2 I$. \QEDB
\end{note}
\begin{lem} \label{lem_t}  
Suppose at the $(k-1)^{th}$ time index ($k\geq 1$), the actuators which successfully receive the control signals are those indexed by the elements of $\mathscr{I}_j$, i.e., $\xi_{k-1}=\mathcal{N}(j)$. Then, for the Isaacs equation (\ref{value_fn1_multi}) the following claims are true:
	\begin{enumerate} [(a)]
		\item The value function  at the stage $k\in [0,N]$ is well defined if and only if:
		\begin{enumerate}[(i)]
			\item
			\small
			\begin{equation}
			\Theta_{k,N}(i,j)>0. 
			\label{saddle_multi_channel_1}
			\end{equation} 
			\normalsize
			\item
			\small
			\begin{equation}
			\begin{split}
			&\hspace{2.6cm} \Theta_{l,N}(e,f)>0;\\
			&  k+1\leq l\leq N+1, \forall e\in \mathcal{D}, \forall f\in \{0,1,...,(2^m-1)\}. 
			\end{split}
			\end{equation}
			\normalsize
			where  $\Theta_{k,N}(i,j)$ for $k\in [0,N]$ is as defined in (\ref{theta}) and $\Theta_{N+1,N}(i,j)$ is as given in Note 3.
		\end{enumerate}
	 \item  The value function is given by:
		\small
		\begin{equation}
		V_{k,N}\Big(x_k,i,N(j)\Big)= x_k^T\Xi_{k,N}(i,j) x_k, 
		\label{value_multi}
		\end{equation}
		\normalsize
			where $\Xi_{p,N}(i,j)$ for all $p\in [0,N]$ is defined in (\ref{care_multi}).
		\item $\Xi_{k,N}(i,j)\geq 0$ for all $k\in [0,N]$, $i\in \mathcal{D}$, $j\in \{0,1,...,(2^m-1)\}$.
	\end{enumerate} 
 \end{lem}
\textit{Proof:} We prove the Lemma using induction.\\
  Using (\ref{value_fn1_multi_1}), one can easily express the value function at the stage $k=N$ as $V_{N,N}\Big(x_{N},i,\mathcal{N}(j)\Big)=x_{N}^T\Xi_{N,N}(i,j)x_{N}$, where $\Xi_{N,N}(i,j)=W\geq 0$ for all $i\in \mathcal{D}$, $j\in \{0,1,...,(2^m-1)\}$. Further, in view of Note \ref{note_aa}, for all  $j\in \{0,1,...,(2^m-1)\}$, $i\in \mathcal{D}$, $\Theta_{N+1,N}(i,j)=\gamma^2I>0$. Hence the lemma is true for the case $k=N$. \\
  Suppose, the lemma is true for all the stages $k\geq p+1$.\\
  The necessary part of statement (a) can be proved in exactly the same way as Theorem {3.2} in \cite{n}. We now prove the sufficiency part.\\
  Assume that $\Theta_{l,N}(i,j)>0$ for $p+1\leq l\leq N$, $\forall i\in \mathcal{D}$, and $\forall j\in \{0,1,...,(2^m-1)\}$. Then, with information set $\mathcal{I}_p$, if $\xi_{p-1}=\mathcal{N}(j)$:
  \small
  \begin{equation}
  \begin{split}
  &\mathbb{E}\Big[V_{p+1,N}(x_{p+1},r_{p+1},\xi_{p})\Big| \mathcal{I}_{p}\Big]\\
  & = \sum_{l=0}^{2^m-1}\mathcal{P}_{p}(l)\Big[\Big(A(i)x_p+B(i)\mathcal{N}(l)u_p+D_1(i)w_p\Big)^T\\
  & \times \mathscr{X}_{p+1,N}(i,l) \Big(A(i)x_p+B(i)\mathcal{N}(l)u_p+D_1(i)w_p\Big)\Big].
  \end{split}
  \label{V_1_(k+1)} 
  \end{equation}  
  \normalsize 
\\ From (\ref{value_fn1_multi}) with $k=p$, and using (\ref{V_1_(k+1)}): 
\small
\begin{equation} 
\begin{split}
&V_{p,N}\Big(x_{p},r_{p},\xi_{p-1}\Big) = \underset {u_{p}}{\textrm{min}} \hspace{0.1cm} \underset {w_{p}} {\textrm{max}} \hspace{0.1cm}\mathbb{E} \Big[x_{p}^TW(i)x_{p}+ u^T_{p}\xi_{p}R(i)\xi_{p}u_{p}\\
&-\gamma^2w_{p}^Tw_{p}+\sum_{l=0}^{2^m-1}\mathcal{P}_{p}(l)\Big[\Big(A(i)x_{p}+B(i)\mathcal{N}(l)u_{p}+D_1(i)w_{p}\Big)^T\\
& \times \mathscr{X}_{p+1,N}(i,l)  \Big(A(i)x_{p}+B(i)\mathcal{N}(l)u_{p}+D_1(i)w_{p} \Big) \Big]. 
\end{split}
\label{eq_V_k}
\end{equation} 
\normalsize
Consider the following functional:
\small
\begin{equation}
\begin{split} 
&H_{p,N}\Big(x_{p},u_{p},w_{p},r_{p}=i,\xi_{p-1}\Big)=  \mathbb{E} \Big[ ||x_{p}||_{W(i)}^2 +  ||u_{p}||_{\xi_{p}R(i)\xi_{p}}^2\\
&-\gamma^2 ||w_{p}||^2 + V_{p+1,N}\Big(x_{p+1},r_{p+1},\xi_{p}\Big)\Big| \mathcal{I}_p\Big].
\label{H_k}
\end{split} 
\end{equation}
\normalsize
Observe that $H_{p,N}(x_{p},u_{p},w_{p},i,\xi_{p-1})$ is quadratic in $u_{p}$, $w_{p}$ and $x_{p}$ for all $i\in \mathcal{D}$. Thus, a unique saddle-point exists if and only if $H_{p,N}\Big(x_{p},u_{p},w_{p},r_{p}=i,\xi_{p-1}\Big)$ is convex in $u_p$ and concave in $w_p$.
Differentiating (\ref{H_k}) and using (\ref{V_1_(k+1)}),
\small
\begin{subequations}
	\begin{equation}
	\begin{split}
	& \frac{\partial^2 H_{p,N}\Big(x_{p},u_{p},w_{p},i,\xi_{p-1}\Big)}{\partial u_{p}^2} =\mathbf{L}_{p}\Big(\mathscr{R}_{p+1,N}(i,.)\Big),
	\end{split}
	\label{H_u}
	\end{equation} 
	\begin{equation}
	\begin{split}
	&\frac{\partial^2 H_{p,N}\Big(x_{p},u_{p},w_{p},i,\xi_{p-1}\Big)}{\partial w_{p}^2}\\
	& =  D_1^T(i)\mathbf{L}_{p}\Big(\mathscr{X}_{p+1,N}(i,.)\Big)D_1(i)-\gamma^2I \\
	&= - \Theta_{p,N}(i,j).
	\end{split}
	\label{H_w}
	\end{equation}
\end{subequations}
\normalsize
By our assumption $\Xi_{p+1,N}(i,j)\geq 0$ for all $i\in \mathcal{D}$, $j\in\{0,1,...,(2^m-1)\}$. Thus $\mathscr{X}_{p+1,N}(i)\geq 0$, $i\in \mathcal{D}$ and $j\in\{0,1,...,(2^m-1)\}$.  So, $R(i)>0$ implies $\mathscr{R}_{p+1,N}(i,j)> 0$. Hence, $ \frac{\partial^2 H_{{p},N}\Big(x_{p},u_{p},w_{p},i,\mathcal{N}(j)\Big)}{\partial u_{p}^2}>0$. Therefore, $H_{{p},N}\big(x_{p},u_{p},w_{p},i,\mathcal{N}(j)\big)$ is convex in $u_{p}$. Also, if $\Theta_{p,N}(i,j)>0$, then $H_{{p},N}\big(x_{p},u_{p},w_{p},i, \mathcal{N}(j)\big)$ will be concave in $w_{p}$.
Now, if $(u_p^*,w_p^*)$ is the saddle-point, then:
\small
\begin{equation}
\begin{split}
&\frac{\partial H_{p,N}\Big(x_{p},u_{p},w_{p},i,\xi_{p-1}\Big)}{\partial u_p}\Big|_{u_p^*,w_p^*}=0 \\
&\frac{\partial H_{p,N}\Big(x_{p},u_{p},w_{p},i,\xi_{p-1}\Big)}{\partial w_p}\Big|_{u_p^*,w_p^*}=0.
\end{split}
\label{saddle_nec_1}
\end{equation}
\normalsize
 Solving (\ref{saddle_nec_1}), one gets:
\small
\begin{subequations}
	\begin{equation}
	\begin{split}
	u_{p}^{*} &=\zeta_{p}^*(\mathcal{I}_{p})=-\Gamma_{p,N}(i,j)x_{p}
	\end{split}
	\label{optimal_u}
	\end{equation} 
	\begin{equation} 
	\begin{split}
	w_{p}^{*}&  = \eta_{p}^*(\mathcal{I}_{p})=\Psi_{p,N}(i,j)x_{p},
	\end{split}
	\label{optimal_w}
	\end{equation} 
\end{subequations}
\normalsize
where, $\Gamma_{p,N}(i,j)$ and $\Psi_{p,N}(i,j)$ are as defined in (\ref{Gamma_multi}) and (\ref{Psi_multi}), respectively. As $\mathscr{R}_{p+1,N}(i,j)> 0$, invertibility of $\Lambda_{p,N}(i,j)$ for all $ i\in \mathcal{D}$ and $j\in \{0,1,...,(2^m-1)\}$ is guaranteed. Further, invertibility of  $\Lambda_{p,N}(i)$ ensures finiteness of $\Psi_{p,N}(i,j)$ if $\Theta_{p,N}(i,j)>0$. Thus, $\Gamma_{{p},N}(i,j)$ is finite.
Substituting the saddle-point $\big(u^*_{p},w^*_{p}\big)$ from (\ref{optimal_u}) and (\ref{optimal_w}) in (\ref{eq_V_k}) with $k=p$: 
\small
\begin{equation*}
V_{p,N}\big(x_{p},i, \mathcal{N}(j)
\big)=x_{p}^T\Xi_{{p},N}(i,j)x_{p},	
\end{equation*}
\normalsize
where $\Xi_{p,N}(i,j)$ is got by solving (\ref{care_multi}).
\\ Observe that $H_{p,N}\big(x_{p},u_{p}^*,w_{p}=0,i,\mathcal{N}(j)\big)\geq 0$ for all $ i\in \mathcal{D}$ and $j\in \{0,1,...,(2^m-1)\}$ as $\Xi_{p+1,N}(i,j)\geq 0$ for all $i\in \mathcal{D}$, $j\in \{0,1,...,(2^m-1)\}$. Also, as $w_{p}$ is the maximizing player: 
\small
\begin{equation*}
\begin{split}
&H_{p,N}\big(x_{p},u_{p}^*,w_{p}^*,i,\mathcal{N}(j)\big)\geq H_{p,N}\big(x_{p},u_{p}^*,0,i,\mathcal{N}(j)\big)\geq 0.
\end{split}
\end{equation*} 
\normalsize
 Therefore, 
\small
\begin{equation}
\begin{split}
V_{p,N}\big(x_{p},i, \mathcal{N}(j)
\big)&=x^T_{p}\Xi_{p,N}(i,j)x_{p}
\\&=H_{p,N}\big(x_{p},u_{p}^*,w_{p}^*,i,\mathcal{N}(j)\big)\\
& \geq 0.
\end{split}
\label{ppp_1}
\end{equation}
\normalsize
 As (\ref{ppp_1}) holds for all $x_{p}\neq 0$, $\Xi_{p,N}(i,j)\geq 0$. Hence, the lemma is true for the stage $k=p$. \QEDB
 %\par \textit{Proof:} We can proof the lemma by using the same line of argument as used in the proof of \textit{Lemma 2.} 
% The following result demonstrate that the finite horizon controller attains requirement (R.1).
\par As a direct consequence of Lemma \ref{lem_t}, we get the following result.
\begin{cor} \label{cor_1}
At the stage $k\in[0,N-1]$, a unique saddle-point exists if and only if conditions (i) and (ii) given in (a) in Lemma 2 are satisfied.
Also, the saddle-point at the stage $k\in [0,N-1]$ is as given by (\ref{optimal_u}) and (\ref{optimal_w}). \QEDB
\end{cor}
\begin{note}
One can observe that for a fixed $r_0$, $\Xi_{0,N}(r_0,0)=\Xi_{0,N}(r_0,1)=...=\Xi_{0,N}\big(r_0,(2^m-1)\big)$. Hence, if a unique saddle-point exists at the stage $k=0$,
 $V_{0,N}\Big(x_0,r_0,\mathcal{N}(0)\Big)=V_{0,N}\Big(x_0,r_0,\mathcal{N}(1)\Big)=...=V_{0,N}\Big(x_0,r_0,\mathcal{N}(2^m-1)\Big)=\hat{V}_{0,N}\Big(x_0,r_0\Big)$ (say). The value of the game with the cost function (\ref{cost_fun_multi_channel}) is then given by:
\small
\begin{equation}
\begin{split}
J_N({\zeta}_{0:N-1}^{*},{\eta}_{0:N-1}^{*}) =\hat{V}_{0,N}\Big(x_0,r_0\Big) = x^T_0\hat{\Xi}_{0,N}(r_0)x_0,
\end{split}  
\label{eq_V_0} 
\end{equation} 
\normalsize
where $\hat{\Xi}_{0,N}(r_0)=\Xi_{0,N}(r_0,j)$ for all $j\in \{0,1,...,(2^m-1)\}$. 
\par \QEDB
\end{note}
\begin{lem} \label{lem_j} Consider that a unique saddle-point exists at the stage $k=0$. Then, with the optimal control sequence $u^*_{0:N-1}$, $\mathcal{L}_2$ gain from disturbance $w_k$ to controlled output $z_k$ of  the closed loop system is less than or equal to $\gamma$.
\end{lem}  
\par \textit{Proof:}
 If a unique saddle-point exists at the stage $k=0$, then,
\small
\begin{equation*}
\begin{split}
&J_N(\zeta^{*N-1},\eta^{N-1})\leq J_N(\zeta^{*N-1},\eta^{*N-1}) \leq J_N(\zeta^{N-1},\eta^{*N-1})\\
&\implies J_N(\zeta^{*N-1},\eta^{N-1})\leq J_N(\zeta^{*N-1},\eta^{*N-1})=x_0^T\hat{\Xi}_{0,N}(r_0)x_0\\
& \hspace{7cm} \Big(\textrm{using} (\ref{eq_V_0})\Big)\\
& \implies J_N(\zeta^{*N-1},\eta^{N-1})\leq 0\\
& \hspace{4cm} \Big(\textrm{considering zero initial condition}\Big)\\
& \implies \mathbb{E} \Big[||x_{N}||_{W}^2+\sum_{k=0}^{N-1} ||z_k||^2-\gamma^2||w_k||^2 \Big| \mathcal{I}_0 \Big] \leq 0\\
& \hspace{7cm} \Big(\textrm{using} (\ref{cost_fun1})\Big)\\
& \implies \sum_{k=0}^{N-1}\mathbb{E}\Big[||z_k||^2\Big| \mathcal{I}_0\Big]\leq \gamma^2 \sum_{k=0}^{N-1} ||w_k||^2.
\end{split}
\label{saddle_condition}
\end{equation*}
\normalsize
\vspace{-0.0cm}
Therefore, $\mathcal{L}_2$ gain from the disturbance input $w_k$ to the controlled output $z_k$ is less than or equal to $\gamma$. It completes the proof.   \QEDB 
\par In the subsequent results, we shall assume that $W(r_k)\geq W$ for all $r_k$. %Lemma \ref{lem_y}, Lemma \ref{lem_pp}, and Lemma \ref{lem_aa} present some important properties of $\Xi_{k,N}(i,j)$. These results will be used in the infinite horizon case.  
\begin{lem} \label{lem_y} If a unique saddle-point exists at the stage $k\in [1,N]$, then $\Xi_{k,N}(i,j)\geq\Xi_{k+1,N}(i,j)$ for all  $k\geq 1$, $i \in \mathcal{D}$ and $j\in\{0,1,...,(2^m-1)\}$.
\end{lem}
\par \textit{Proof:}
This result is proved using induction. \\
From Note \ref{note_aa}, $\Xi_{N+1,N}(i,j)=0$. Thus:
\begin{equation*}
\begin{split}
\Xi_{N,N}(i,j)=W\geq 0 =\Xi_{N+1,N}(i,j).
\end{split}
\end{equation*}
For the stage $k=p$, from (\ref{eq_V_k}):
\small
\begin{equation}
\begin{split}
& V_{p,N}\Big(x,i,\mathcal{N}(j)\Big) \\
&= x^T\Xi_{p,N}(i,j)x\\
&=\underset {u} {\textrm{min}} \hspace{0.2cm} \underset {w} {\textrm{max}}  \Big[x^TW(i)x+u^T\mathbf{L}_p\Big(\mathscr{Q}(i,.)\Big)u-\gamma^2w^Tw\\
&+ \sum_{l=0}^{2^m-1}\mathcal{P}_p(l) \Big(A(i)x+B(i)\mathcal{N}(l)u+D_1(i)w\Big)^T \mathscr{X}_{p+1,N}(i,.)\\
& \times \Big(A(i)x+B(i)\mathcal{N}(l)u+D_1(i)w\Big)\Big].
\end{split}
\label{mono_1}
\end{equation} 
\normalsize
Replacing $p$ by $p+1$ in the above equation:
\small
\begin{equation}
\begin{split}
& V_{p+1,N}\Big(x,i,\mathcal{N}(j)\Big) =x^T\Xi_{p+1,N}(i,j)x\\
& = \underset {u} {\textrm{min}} \hspace{0.2cm} \underset {w} {\textrm{max}}  \Big[x^TW(i)x+u^T\mathbf{L}_{p+1}\Big(\mathscr{Q}(i,.)\Big)u-\gamma^2w^Tw\\
&+ \sum_{l=0}^{2^m-1}\mathcal{P}_{p+1}(l) \Big(A(i)x+B(i)\mathcal{N}(l)u+D_1(i)w\Big)^T \mathscr{X}_{p+2,N}(i,.)\\
& \times \Big(A(i)x+B(i)\mathcal{N}(l)u+D_1(i)w\Big) \Big].
\end{split}
\label{mono_2}
\end{equation}
\normalsize
Suppose, the lemma is true for the stage $k=p+1$. Thus, $\Xi_{p+1,N}(i,j) \geq \Xi_{p+2,N}(i,j)$ for all $i\in \mathcal{D}$ and $j\in \mathcal{N}(j)$. Note that, $\mathcal{P}_p(.)=\mathcal{P}_{p+1}(.)$ and $\mathbf{L}_p(.)=\mathbf{L}_{p+1}(.)$ if $\xi_{p-1}=\xi_{p}$. Further, for two functions $g_1(u,w)$ and $g_2(u,w)$, if $g_1(u,w)\geq g_2(u,w)$ for all $u,w$, then one can show that:
$\underset{u}{\textrm{min}}\hspace{0.1cm}\underset{w}{\textrm{max}} \hspace{0.2cm} g_1(u,w)\geq \underset{u}{\textrm{min}}\hspace{0.1cm}\underset{w}{\textrm{max}} \hspace{0.2cm} g_2(u,w). $ 
\normalsize
%\vspace{-0.3cm}
 Then, from (\ref{mono_1}) and (\ref{mono_2}):
\small
\begin{equation}
\begin{split}
&V_{p,N}\Big(x,i,\mathcal{N}(j)\Big)\geq V_{p+1,N}\Big(x,i,\mathcal{N}(j)\Big)\\
&\implies x^T\Xi_{p,N}(i,j)x\geq x^T\Xi_{p+1,N}(i,j)x
\end{split}
\label{ineq}
\end{equation} 
\normalsize
Since (\ref{ineq}) holds true for all $x\neq 0$, $\Xi_{p,N}(i,j) \geq \Xi_{p+1,N}(i,j)$ for all $i\in \mathcal{D}$ and $j\in \{0,1,...,(2^m-1)\}$. \QEDB
\begin{lem} \label{lem_pp} Suppose a unique saddle-point exists for the stage $k\geq 1$. Then, $\Xi_{k,N}(i,j)=\Xi_{k+1,N+1}(i,j)$ for all $k\geq 1$, $i\in\mathcal{D}$ and $j\in \{0,1,...,(2^m-1)\}$.
	\par \textit{Proof:} To prove the lemma, we again use induction.\\
	For the base case, one gets that: $V_{N,N}\Big(x,i,\mathcal{N}(j)\Big)=V_{N+1,N+1}\Big(x,i,\mathcal{N}(j)\Big)=x^TWx$. \\
	Consider the following:
	\small
	\begin{equation}
	\begin{split}
	&V_{p,N}\Big(x,i,\mathcal{N}(j)\Big) =x^T\Xi_{p,N}(i,j)x\\
	& = \underset {u} {\textrm{min}} \hspace{0.2cm} \underset {w} {\textrm{max}}  \Big[x^TW(i)x+u^T\mathbf{L}_{p}\Big(\mathscr{Q}(i,.)\Big)u-\gamma^2w^Tw\\
	&+ \sum_{l=0}^{2^m-1}\mathcal{P}_{p}(l) \Big(A(i)x+B(i)\mathcal{N}(l)u+D_1(i)w\Big)^T \mathscr{X}_{p+1,N}(i,.)\\
	& \times \Big(A(i)x+B(i)\mathcal{N}(l)u+D_1(i)w\Big)\Big]
	\end{split}
	\label{eq_v}
	\end{equation}
	\begin{equation}
	\begin{split}
	&V_{p+1,N+1}\Big(x,i,\mathcal{N}(j)\Big)\\
	& =x^T\Xi_{p+1,N+1}(i,j)x.\\
	&= \underset {u} {\textrm{min}} \hspace{0.2cm} \underset {w} {\textrm{max}}  \Big[x^TW(i)x+u^T\mathbf{L}_{p+1}\Big(\mathscr{Q}(i,.)\Big)u-\gamma^2w^Tw\\
	&+ \sum_{l=0}^{2^m-1}\mathcal{P}_{p+1}(l) \Big(A(i)x+B(i)\mathcal{N}(l)u+D_1(i)w\Big)^T \\
	& \times \mathscr{X}_{p+2,N+1}(i,.) \Big(A(i)x+B(i)\mathcal{N}(l)u+D_1(i)w\Big)\Big]
	\end{split}
	\label{eq_v_2}
	\end{equation}
	\normalsize
	Assume now that the lemma is true for the stage $k=p+1$. Hence, $\Xi_{p+1,N}(i,j)=\Xi_{p+2,N+1}(i,j)$,  $\forall i\in\mathcal{D}$ and $\forall j\in \{0,1,...,(2^m-1)\}$. Therefore, $\mathscr{X}_{p+1,N}(i,j)=\mathscr{X}_{p+2,N+1}(i,j)$. Also,  $\mathcal{P}_p(.)=\mathcal{P}_{p+1}(.)$ and $\mathbf{L}_p(.)=\mathbf{L}_{p+1}(.)$ if $\xi_{p-1}=\xi_{p}$. Therefore, from (\ref{eq_v}) and (\ref{eq_v_2}), one gets that $\Xi_{p,N}(i,j)=\Xi_{p+1,N+1}(i,j)$ for all $i\in \mathcal{D}$ and $j\in \{0,1,...,(2^m-1)\}$.  \QEDB
\end{lem}
\begin{note} \label{note_4}
	From Lemma 4 and Lemma 5, $\Xi_{k,N}(i,j)\geq \Xi_{k+1,N}(i,j)=\Xi_{k,N-1}(i,j)$ or $\Xi_{k,N-1}(i,j) \leq \Xi_{k,N}(i,j)$. Thus, for all $i\in \mathcal{D}$ and $j\in \{0,1,...,(2^m-1)\}$, the sequence $\Big{\{}\Xi_{k,c}(i,j)\Big{\}}_{c=k+1}^N:=\Big{\{}\Xi_{k,k+1}(i,j),\Xi_{k,k+2}(i,j),...,\Xi_{k,N}(i,j)\Big{\}}$ monotonically increases with $c$. \QEDB
\end{note}
\begin{lem} \label{lem_aa} Suppose a unique saddle-point exists at the stage $k=0$. Then, the sequence $\Big{\{}\hat{\Xi}_{0,c}(i)\Big{\}}_{c=1}^N:=\Big{\{}\hat{\Xi}_{0,1}(i,j),\hat{\Xi}_{0,2}(i,j),...,\hat{\Xi}_{0,N}(i,j)\Big{\}}$ monotonically increases with $c$.
	\par \textit{Proof:} Using (\ref{value_fn1_multi}), for a horizon $ l\in [1,N]$ one can write the value function at the stage $k=0$ as:
	\small
	\begin{equation}
	\begin{split}
	V_{0,l}(x_0,i)=\mathbb{E}\Big[x_0^TW(i)x_0+u_0^T\hat{\mathbf{L}}\Big(\mathscr{Q}(i,.)\Big)u_0-\gamma^2w_0^Tw_0\\
	+V_{1,l}\Big(x_1,r_1,\xi_0\Big) \Big| \mathcal{I}_0 \Big].
	\end{split}
	\label{ppp}
	\end{equation}
	\normalsize
	 Considering the horizon to be $l+1$, we can write (\ref{ppp}) as:
	 \small
	 \begin{equation}
	 \begin{split}
	 V_{0,l+1}(x_0,i)=\mathbb{E}\Big[x_0^TW(i)x_0+u_0^T\hat{\mathbf{L}}\Big(\mathscr{Q}(i,.)\Big)u_0-\gamma^2w_0^Tw_0\\
	 +V_{1,l+1}\Big(x_1,r_1,\xi_0\Big) \Big| \mathcal{I}_0\Big].
	 \end{split}
	 \label{ppp_2}
	 \end{equation}
	 \normalsize
	From Note \ref{note_4},
	\begin{equation*}
	\mathbb{E}\Big[V_{1,l}\Big(x_1,r_1,\xi_0\Big)\Big| \mathcal{I}_0\Big]\leq \mathbb{E}\Big[V_{1,l+1}\Big(x_1,r_1,\xi_0\Big)\Big| \mathcal{I}_0\Big].
	\end{equation*} 
	 Thus from (\ref{ppp}) and (\ref{ppp_2}): 
	 \small
	 \begin{equation}
	 \begin{split}
	  V_{0,l}(x_0,i)\leq V_{0,l+1}(x_0,i)\\
	  \implies x_0^T\hat{\Xi}_{0,l}(i)x_0\leq x_0^T\hat{\Xi}_{0,l+1}(i)x_0.
	 \end{split}
	 \label{erd}
	 \end{equation}
	 \normalsize
\end{lem}
\normalsize
Since (\ref{erd}) is satisfied for all $x_0\neq 0$, $\hat{\Xi}_{0,l}(i)\leq \hat{\Xi}_{0,l+1}(i)$ for all $i\in \mathcal{D}$. \QEDB 
\vspace{-0.3cm}
\subsection{{Infinite Horizon Control:}}
For the infinite horizon case, we consider the following cost function:
\small
\begin{equation}
\begin{split}
J_\infty(\zeta_{0:\infty},\eta_{0:\infty})=\mathbb{E} \Big[\sum_{k=0}^{\infty} x_k^TW(r_k)x_k+u_k^T \xi_k^TR(r_k)\xi_ku_k\\-\gamma^2w_k^Tw_k \Big| \mathcal{I}_0\Big] .
\end{split}
\label{cost_fun_multi_channel_inf}
\end{equation}
\normalsize
In the next lemma, we provide certain conditions for convergence of the sequence $\Big{\{}\Xi_{c,N}(i,j)\Big{\}}_{c=k+1}^N$ as $N\rightarrow \infty$.
\begin{lem} \label{lem_vv} Suppose $\gamma$ and $\bar{v}^{1:m}$, $\bar{\mu}^{1:m}$ are chosen such that for all finite $N\in \mathbb{Z}^+$, $k\in[0,N]$, $i\in \mathcal{D}$, $j\in \{0,1,...,(2^m-1)\}$,
	\begin{enumerate}[(i)]
	\item 	$\Theta_{k,N}(i,j)>0$.
	\item  $\Xi_{k,N}(i,j)<\infty$.
	\end{enumerate}
	  Then, there exist $\bar{\hat{\Xi}}(i)<\infty$ and $\bar{\Xi}(i,j)<\infty$ such that $\underset{N\rightarrow \infty}{\textrm{lim}}\hat{\Xi}_{0,N}(i)= \bar{\hat{\Xi}}(i)$, and for $k\geq 1$, $\underset{N\rightarrow \infty}{\textrm{lim}}\Xi_{k,N}(i,j)= \bar{\Xi}(i,j)$ for all $i\in \mathcal{D}$, $j\in \{0,1,2,...,(2^m-1)\}$.
\end{lem}
\par \textit{Proof:} From Note \ref{note_4} and Lemma \ref{lem_aa}, monotonicity of the sequence $\Big{\{}\Xi_{k,c}(i,j)\Big{\}}_{c=k+1}^N$ for $k\in[0,\infty)$ is guaranteed if condition (i) is satisfied. Further, suppose $\gamma$ and $\bar{v}^{1:m}$, $\bar{\mu}^{1:m}$ are chosen such that for all finite $N\in \mathbb{Z}^+$, $k\in[0,N]$, $i\in \mathcal{D}$, $j\in \{0,1,...,(2^m-1)\}$ we have  $\Xi_{k,N}(i,j)<\infty$. Hence, in view of {Theorem 3.14} in \cite{rudin1976principles}, $\Big{\{}\hat{\Xi}_{0,N}(i)\Big{\}}_{c=1}^N$, and  $\Big{\{}\Xi_{k,N}(i,j)\Big{\}}_{c=k+1}^N$ for all $k\in[1,N-1]$ converge as $N\rightarrow \infty$. Therefore, there exist $\bar{\hat{\Xi}}(i)<\infty$ and $\bar{{\Xi}}(i,j)<\infty$ such that $\underset{N\rightarrow \infty}{\textrm{lim}}\hat{\Xi}_{0,N}(i)= \bar{\hat{\Xi}}(i)$, and for $k\in[1,N]$, $\underset{N\rightarrow \infty}{\textrm{lim}}\Xi_{k,N}(i,j)= \bar{\Xi}(i,j)$ for all $i\in \mathcal{D}$, $j\in \{0,1,2,...,(2^m-1)\}$. 
 \QEDB
\begin{remark}
	For the case when the matrices $D_1(i)$, $i\in \mathcal{D}$ in (\ref{system_eq}) are square and full rank, $\Theta_{k,N}(i,j)>0$ for all $i\in \mathcal{D}$, $j\in \{0,1,...,(2^m-1)\}$ ensures that $\Xi_{k,N}(i,j)<\infty$. In such a scenario, condition (i) in Lemma (\ref{lem_vv}) alone is sufficient for convergence of the sequence $\Big{\{}\Xi_{k,c}(i,j)\Big{\}}_{c=k+1}^N$ as $N\rightarrow \infty$. 
	\par \QEDB
\end{remark}
\par If for all $k \in [1,N-1]$, $i\in \mathcal{D}$, $j\in \{0,1,..,(2^m-1)\}$, the sequence $\Big{\{}\Xi_{k,c}(i,j)\Big{\}}_{c=k+1}^N$ converges as $N\rightarrow \infty$, $\Xi_{k,N}(i,j)$, $\Gamma_{k,N}(i,j)$, $\Theta_{k,N}(i,j)$, etc., will no longer be functions of $k$. The CAREs (\ref{care_multi}) then transform to the following: 
\small
\begin{equation}
\begin{split}
& \bar{\Xi}(i,j)= W(i)+\bar{\Gamma}^T(i,j)\mathbf{L}^j\Big(\mathscr{Q}(i,.)\Big)\bar{\Gamma}(i,j)-\gamma^2\bar{\Psi}^T(i,j)\bar{\Psi}(i,j)\\
&  +\sum_{l=0}^{2^m-1}{\mathcal{P}}^j(l)\Big[\Big(A(i)-B(i)\mathcal{N}(l)\bar{\Gamma}(i,j)+D_1(i)\bar{\Psi}(i,j)\Big)^T\bar{\mathscr{X}}(i,l)\\
& \times \Big(A(i)-B(i)\mathcal{N}(l)\bar{\Gamma}(i,j)+D_1(i)\bar{\Psi}(i,j)\Big)\Big],
\end{split}
\label{care_multi_infi}
\end{equation}
\normalsize
where $\bar{\Gamma}(i,j)$, $\bar{\Psi}(i,j)$ are the infinite-horizon counterparts of $\Gamma_{k,N}(i,j)$ and $\Psi_{k,N}(i,j)$. Replacing $\Xi_{k,N}(i,j)$ by $\bar{\Xi}(i,j)$ in (\ref{Gamma_multi}) and (\ref{Psi_multi}), $\bar{\Gamma}(i,j)$ and $\bar{\Psi}(i,j)$ can be derived easily.
\normalsize
\par
Similarly, for the stage $k=0$, infinite horizon CAREs takes the following form:
 \small
 \begin{equation}
 \begin{split}
 & \bar{\hat{\Xi}}(i)= W(i)+\bar{\hat{\Gamma}}^T(i)\hat{\mathbf{L}}\Big(\mathscr{Q}(i,.)\Big)\bar{\hat{\Gamma}}(i)-\gamma^2\bar{\hat{\Psi}}^T(i)\bar{\hat{\Psi}}(i)\\
 &  +\sum_{l=0}^{2^m-1}{\hat{\mathcal{P}}}(l)\Big[\Big(A(i)-B(i)\mathcal{N}(l)\bar{\hat{\Gamma}}(i)+D_1(i)\bar{\hat{\Psi}}(i)\Big)^T\bar{\mathscr{X}}(i,l)\\
 & \times \Big(A(i)-B(i)\mathcal{N}(l)\bar{\hat{\Gamma}}(i)+D_1(i)\bar{\hat{\Psi}}(i)\Big)\Big],
 \end{split}
 \label{care_multi_infi_0}
 \end{equation}
 \normalsize
 where $\bar{\hat{\Gamma}}(i)$ and $\bar{\hat{\Psi}}(i)$ can be got from  (\ref{Gamma_multi}) and (\ref{Psi_multi}) by replacing $\mathbf{L}_k(.)$ by $\hat{\mathbf{L}}(.)$ and $\Xi_{k,N}(i)$ by $\bar{\hat{\Xi}}(i)$. 
 \normalsize
Further, the infinite horizon Isaacs Equation is as given below:
\small
\begin{equation}
\begin{split}
 V(x_k,r_k,\xi_{k-1}) = \underset {u_k}{\textrm{min}} \hspace{0.1cm} \underset {w_k} {\textrm{max}} \hspace{0.1cm}\mathbb{E} \hspace{0.1cm} \Big[ x_k^TW(r_k)x_k+u_k^T \xi_k^TR(r_k)\xi_ku_k\\ -\gamma^2w_k^Tw_k+V(x_{k+1},r_{k+1},\xi_k)|\mathcal{I}_k\Big].
\end{split}
\label{value_fn1_infi_multi}
\end{equation}
\normalsize
One gets the following result by applying limit $N\rightarrow \infty$ in Lemma \ref{lem_t}. 
\begin{lem} \label{lem_u}  Suppose $\bar{v}^{1:m}$, $\bar{\mu}^{1:m}$ and $\gamma$ are chosen such that for all finite  $N\in \mathbb{Z}^+$, $k\in [0,N]$,  $ i\in\mathcal{D}$, $ j\in \{0,1,...,(2^m-1)\}$, conditions (i) and (ii) given in the {Lemma \ref{lem_vv}} are satisfied. Then,
 \begin{enumerate}[(a)]
 	\item The value function at the stage $k\in [1,\infty]$ is expressed as:
 	\small
 	\begin{equation*}
 	V\Big(x_k,i,N(j)\Big)= x_k^T\bar{\Xi}(i,j) x_k.
 	\end{equation*}
 	\normalsize
 	\item The value function at the $k=0$ stage is expressed as:
 	\small
 	\begin{equation*}
 	V\Big(x_{0},r_0\Big)
 	=x_0^T\bar{\hat{\Xi}}(r_0)x_0.
 	\end{equation*}
 	\normalsize
 	\item The infinite horizon saddle-point is given by:
 	\small
 	\begin{equation*}
 	u_k^*=-\tilde{\Gamma}_k x_k;\hspace{0.2cm} w_k^*=\tilde{\Psi}_kx_k,
 	\end{equation*}
 	\normalsize
 	where
 	\small
 	\begin{equation}
 	\begin{split}
 	&\textrm{For} \hspace{0.2cm} {k\geq 1} \hspace{0.2cm}, \hspace{0.2cm} \textrm{if} \hspace{0.2cm} r_k=i\hspace{0.2cm} \textrm{and} \hspace{0.2cm} \xi_{k-1}=\mathcal{N}(j), \\ &\textrm{then}, \hspace{0.2cm} \tilde{\Gamma}_k=\bar{\Gamma}(i,j), \hspace{0.2cm} \tilde{\Psi}_k=\bar{\Psi}(i,j).\\
 	& \textrm{For} \hspace{0.2cm} k=0, \hspace{0.2cm} \textrm{if} \hspace{0.2cm} r_0=i, \\ 
 	&\textrm{then}, \hspace{0.2cm} \tilde{\Gamma}_0=\bar{\hat{\Gamma}}(i), \hspace{0.2cm} \tilde{\Psi}_0=\bar{\hat{\Psi}}(i).   
 	\end{split}
 	\label{g_f}
 	\end{equation}
 	\normalsize
 	\item The infinite horizon value of the game with the cost function (\ref{cost_fun_multi_channel_inf}) is expressed as follows:
 	\small
 	\begin{equation*}
 	\begin{split}
 	J_{\infty}(\zeta^*_{0:\infty},\eta^*_{0:\infty})
 	=x_0^T\bar{\hat{\Xi}}(r_0)x_0. 
 	\normalsize \hspace{3.7cm} \QEDB
 	\end{split}
 	\end{equation*} 
 \end{enumerate} 
\end{lem}
\vspace{-0.3cm}
The following lemma establishes the positive definiteness of the fixed-point solutions of CAREs (\ref{care_multi_infi}) and (\ref{care_multi_infi_0}).
\begin{lem} \label{lem_q_multi}Suppose, $\gamma$, $\bar{v}^{1:m}$ and $\bar{\mu}^{1:m}$ are chosen such that for all $i\in \mathcal{D}$, $j\in \{0,1,...,(2^m-1)\}$, and finite $N\in \mathbb{Z}^+$, $k\in[0,N]$, the conditions (i) and (ii) of {Lemma \ref{lem_vv}} are satisfied. Further, if system (\ref{system_eq}) with $u_k\equiv 0$, $w_k\equiv 0$ is weakly observable, then, for all $ i\in \mathcal{D}$, $j\in \{0,1,...,(2^m-1)\}$, $\bar{\hat{\Xi}}(i)>0$ and $\bar{\Xi}(i,j)>0$. 
\end{lem}
\par \textit{Proof:} Consider the following functional:
\small
\begin{equation}
\begin{split}
H_{k,N}\Big(x_{k},u_{k},w_{k},i,\xi_{k-1}\Big)=  \mathbb{E} \Big[ ||x_{k}||_{W(i)}^2 +  ||u_{k}||_{\mathbf{L}_{k}\Big(\mathscr{Q}(i,.)\Big)}^2\\
-\gamma^2 ||w_{k}||^2+ V_{k+1,N}\Big(x_{k+1},r_{k+1},\xi_k\Big) \Big| \mathcal{I}_k\Big].
\end{split}
\label{H_k_k_k}
\end{equation}
\normalsize
\vspace{-0.2cm}
As $(u_k^*,w_k^*)$ constitutes a saddle-point, 
\small
\begin{equation*}
H_{k,N}\Big(x_{k},u^*_{k},w^*_{k},i,\xi_{k-1}\Big)\geq H_{k,N}\Big(x_{k},u^*_{k},0,i,\xi_{k-1}\Big).
\end{equation*} 
\normalsize
Taking limit as $N\rightarrow \infty$:
\small
\begin{equation}
\underset{N\rightarrow \infty}{\textrm{lim}} H_{k,N}\Big(x_{k},u^*_{k},w^*_{k},i,\xi_{k-1}\Big)\geq \underset{N\rightarrow \infty}{\textrm{lim}} H_{k,N}\Big(x_{k},u^*_{k},0,i,\xi_{k-1}\Big).
\label{H_k_inf}
\end{equation} 
\normalsize
Observe that
\small
\begin{equation}
\begin{split}
& \underset{N\rightarrow \infty}{\textrm{lim}} H_{k,N}\Big(x_{k},u^*_{k},w^*_{k},i,\xi_{k-1}=\mathcal{N}(j)\Big)\\
&=  V\Big(x_k,i,\mathcal{N}(j)\Big) \\
&= \underset{N\rightarrow \infty}{\textrm{lim}} \sum_{f=k}^{N} \mathbb{E} \Big[ x_f^TW(i)x_f+ x_f^T\tilde{\Gamma}^T_{f} \mathbf{L}_{f}\big(\mathscr{Q}(i,.)\big)\tilde{\Gamma}_fx_f\\
&-\gamma^2w_f^{*T}w_f^*\Big| \mathcal{I}_k\Big]. \hspace{0.2cm} \Big(\textrm{using (\ref{value_fn1_multi_1})}\Big)
\end{split}
\label{rrr}
\end{equation}
\normalsize
Also 
\small
\begin{equation}
\begin{split}
& \underset{N\rightarrow \infty}{\textrm{lim}} H_{k,N}\Big(x_{k},u^*_{k},0,i,\xi_{k-1}=\mathcal{N}(j)\Big)\\
&= \underset{N\rightarrow \infty}{\textrm{lim}} \sum_{f=k}^{N} \mathbb{E} \Big[ x_f^TW(i)x_f+ x_f^T\tilde{\Gamma}^T_{f} \mathbf{L}_{f}\big(\mathscr{Q}(i,.)\big)\tilde{\Gamma}_fx_f\Big| \mathcal{I}_k\Big].
\end{split}
\label{rrr_1}
\end{equation}
\normalsize
Suppose $\tilde{\Gamma}_fx_f\neq 0$ for any $r_f\in \mathcal{D}$ \big(see the definition of $\tilde{\Gamma}_f$ in (\ref{g_f})\big). Then as $\mathbf{L}_{f}\big(\mathscr{Q}(i,.)\big) >0$ for all $f$, and $W(r_f)\geq 0$ for all $r_f\in \mathcal{D}$, one gets: 
\small
\begin{equation*}
H_{k,N}\Big(x_{k},u^*_{k},0,i,\xi_{k-1}=\mathcal{N}(j)\Big) > x_k^TW(i)x_k.
\end{equation*}
\normalsize
Now, consider the case when $\tilde{\Gamma}_fx_f=0$ for all $r_f \in \mathcal{D}$. Then, for $w_k\equiv 0$, the state equation (\ref{system_eq}) for all $k$ transforms into:
\small
\begin{equation}
x_{k+1}=A(r_k)x_k.
\label{state}
\end{equation}
\normalsize 
Therefore, from (\ref{rrr_1}) and (\ref{state}):
\small
\begin{equation}
\begin{split}
&\underset{N\rightarrow \infty}{\textrm{lim}} H_{k,N}\Big(x_{k},u^*_{k},0,i,\xi_{k-1}=\mathcal{N}(j)\Big)\\
&= \underset{N\rightarrow \infty}{\textrm{lim}}  \sum_{f=k}^{N} \mathbb{E} \Big[  x_f^TW(r_f)x_f\Big| \mathcal{I}_k\Big]\\
&= \mathbb{E} \Big[ x_k^TW(i)x_k+x_{k+1}^TW(r_{k+1})x_{k+1}\\
&+.....+x_{p}^TW(r_{p})x_{p}+x_{p}^TA^T(r_{p})W(r_{p+1})A(r_{p})x_{p}+...\\
&+x_{p}^T\Big(\Pi_{l=p}^{{p+\mathscr{F}}}A(r_l)\Big)^TW(r_{p+\mathscr{F}})\Big(\Pi_{l=p}^{{p+\mathscr{F}}}A(r_l)\Big)x_{p}\\
&+\underset{N\rightarrow \infty}{\textrm{lim}} \vspace{-0.4cm} \sum_{t=p+\mathscr{F}+1}^{N}x_t^TW(r_{t})x_t\Big| \mathcal{I}_k\Big]\\
&=\mathbb{E}\Big[x_k^TW(i)x_k+x_{k+1}^TW(r_{k+1})x_{k+1}+....\\
&+ x_{p}^T \mathscr{Y}\Big(r_p,r_{p+1},...,r_{p+\mathscr{F}}\Big)x_{p}+\underset{N\rightarrow \infty}{\textrm{lim}} \sum_{t=p+\mathscr{F}+1}^{N} \big[ x_t^TW(t)x_t\big]\Big| \mathcal{I}_k\Big],
\end{split}
\label{H_positive}
\end{equation}
\normalsize
where
\small
\begin{equation*}
\begin{split}
&\mathscr{Y}\Big(r_p,r_{p+1},...,r_{p+\mathscr{F}}\Big)\\
&=\mathscr{O}^T\Big(r_p,r_{p+1},...,r_{p+\mathscr{F}}\Big)\mathscr{O}\Big(r_p,r_{p+1},...,r_{p+\mathscr{F}}\Big).
\end{split}
\end{equation*}
\normalsize
Suppose, for a given Markov chain state $r_{k-1}$, $\mathcal{S}_{k:\mathscr{T}}^{r_k}$ is the set of all transition paths of length $(\mathscr{T}-k)$ which the Markov chain $\{r_k\}$ follows with nonzero probability. Let  $\mathscr{J}(r_{k},...,r_{k+\mathscr{T}}\Big| r_{k-1})$ be the probability that, from the stage $k$ to $k+\mathscr{T}$, the Markov chain $\{r_k\}$ follows a transition path $\{r_{k},r_{k+1},...,r_{k+\mathscr{T}}\}$ given $r_{k-1}$, and $\mathscr{J}(r_{p},...,r_{p+\mathscr{F}}\Big| \{r_{k},...,r_{p-1}\})$ be the probability  that, from the stage $p$ to $p+\mathscr{F}$, the Markov chain $\{r_k\}$ follows a transition path $\{r_{p},r_{p+1},...,r_{p+\mathscr{F}}\}$ given that it has followed the transition path $\{r_{k},r_{k+1},...,r_{p-1}\}$ from the stage $k$ to $p-1$.
 Then,
 \small
 \begin{equation}
 \begin{split}
 & \mathbb{E}\Big{[} x_p^T \mathscr{Y} \Big( r_p,r_{p+1},...,r_{p+\mathscr{F}}\Big) x_p \Big| \mathcal{I}_k
 \Big{]} \\
 =& \sum_{(r_{k+1},...,r_{p-1})\in \mathcal{S}_{k+1:p-1}^{r_k}} \Big[ \mathscr{J}(r_{k+1},...,r_{p-1}\Big| r_k) \Big(\Pi_{l=k}^{p-1} A(r_l)x_k\Big)^T\\
 & \times  \sum_{(r_{p},...,r_{p+\mathscr{F}-1})\in \mathcal{S}_{p:p+\mathscr{F}-1}^{r_{p-1}}}\big[ \mathscr{J}(r_{p},...,r_{p+\mathscr{F}-1}\Big| \{r_{k+1},...,r_{p-1}\})\\
 & \times \mathscr{Y} \Big( r_p,...,r_{p+\mathscr{F}}\Big)  \big]  \Big(\Pi_{l=k}^{p-1} A(r_l)x_k\Big)\Big].
 \label{stab_rt_2}
 \end{split}
 \end{equation}
 \normalsize
 By our assumption, system (\ref{system_eq}) with $u_p\equiv 0$, $w_p\equiv 0$ is weakly observable. Thus, there exists a transition path of finite length $\{i_p,i_{p+1},...,i_{p+\mathscr{F}}\}$ such that the jump observability matrix with respect to that particular transition path has full column rank. Due to irreducibility of the Markov chain $\{r_k\}$, one can choose a finite $p$ such that the probability of occurring such a transition path is nonzero. Hence, the probablity that the matrix $\mathscr{Y} \Big(i_p,i_{p+1},...,i_{p+\mathscr{F}}\Big)$ has full rank would also be nonzero. Therefore, from (\ref{stab_rt_2}):
 \small
 \begin{equation}
 \mathbb{E}\Big{[} x_p^T \mathscr{Y} \Big( r_p,r_{p+1},...,r_{p+\mathscr{F}}\Big) x_p \Big| \mathcal{I}_k
 \Big{]} >0
 \label{stab_rt_1}
 \end{equation}
 \normalsize
Thus, as $W(r_k)\geq 0$ for all $r_k\in \mathcal{D}$, from (\ref{H_positive}) and (\ref{stab_rt_1}):
\small
\begin{equation}
\underset{N\rightarrow \infty}{\textrm{lim}} H_{k,N}\Big(x_{k},u^*_{k},0,i,\xi_{k-1}\Big) > x_k^TW(i)x_k.
\label{ddd}
\end{equation}
\normalsize
Hence, from (\ref{H_k_inf}), (\ref{rrr}) and (\ref{ddd}):
\small
\begin{equation}
\begin{split}
 V\Big(x_k,i,\mathcal{N}(j)\Big)&= x_k^T\tilde{\Xi}_kx_k \hspace{0.3cm} \Big(\textrm{using Lemma \ref{lem_u}}\Big)\\
& =\underset{N\rightarrow \infty}{\textrm{lim}} H_{k,N}\Big(x_k,u_k^*,w_k^*,i,\mathcal{N}(j)\Big)\\
&\geq \underset{N\rightarrow \infty}{\textrm{lim}} H_{k,N}\Big(x_k,u_k^*,0,i,\mathcal{N}(j)\Big) \\
& \Big(\textrm{as $w_k$ is the maximizing player}\Big)\\
& > x_k^TW(i)x_k,
\end{split}
\label{pyy}
\end{equation}
\normalsize
where, for $k\geq 1$, $\tilde{\Xi}_{k}=\bar{\Xi}(i,j)$ if $r_k=i$, $\xi_{k-1}=\mathcal{N}(j)$, and $\tilde{\Xi}_0=\bar{\hat{\Xi}}(i)$ if $r_0=i$.\\
Since (\ref{pyy}) is true for all $x_k\neq 0$, $\tilde{\Xi}_k>0$. Therefore, $\bar{\hat{\Xi}}(i)>0$ and $\bar{\Xi}(i,j)>0$ for all $i\in \mathcal{D}$ and $j\in\{0,1,...,(2^m-1)\}$. \QEDB 
\begin{prop}\label{prop_1}
 \cite{iosifescu2014finite}	An irreducible Markov chain with a finite number of states is always recurrent. \QEDB
\end{prop}
\begin{prop}\label{prop_2}
 \cite{ross2014introduction} For an irreducible, recurrent and aperiodic Markov chain, the limiting distribution for each state is nonzero, i.e.,
\begin{equation*}
\underset{n\rightarrow \infty}{\textrm{lim}} \hspace{0.1cm} Pr\Big(r_n=j\Big| r_0=i\Big)>0, \hspace{0.3cm} \forall i,j\in \mathcal{D}.
\end{equation*} \QEDB
\end{prop}
\par In the following result, we shall show that the optimal controller stabilizes the closed-loop system while maintaining a prescribed $\mathcal{L}_2$ gain.  
  \begin{thm}
  	 Suppose  $\bar{v}^{1:m}$, $\bar{\mu}^{1:m}$ and $\gamma$ are chosen such that  $\forall i\in\mathcal{D}$ and $\forall j\in \{0,1,...,(2^m-1)\}$, and for all finite $N\in \mathbb{Z}^+$, $k\in[0,N]$, the conditions (i) and (ii) of {Lemma \ref{lem_vv}} are satisfied. Further, if system (\ref{system_eq}) with $u_k\equiv 0$ and $w_k\equiv 0$ is weakly observable, then: 
  \begin{enumerate}[(a)]
  \item With the optimal control law ${u}^*_{0:\infty}$, $\mathcal{L}_2$ gain from the disturbance input $w_k$ to the controlled output $z_k$ of the closed loop system is less than or equal to $\gamma$.
  \item The optimal control law ${u}^*_{0:\infty}=-\tilde{{\Gamma}}_k x_k$  stabilizes the system (\ref{system_eq}) with arbitrary disturbance $w_k\in \mathcal{L}_2([0,\infty),\mathbb{R}^s)$.  
  \end{enumerate} 
  \end{thm}
\par \textit{Proof:} Proof for claim (a) follows the same line of argument as the proof for {Lemma \ref{lem_j}}.
\par Since the conditions (i) and (ii) of {Lemma \ref{lem_vv}} are satisfied for all finite $N\in \mathbb{Z}^+$, $k\in[0,N]$,
\small
\begin{equation*}
\begin{split}
& J_{\infty}(\zeta_{0:\infty}^{*},\eta_{0:\infty}) \leq J_{\infty}(\zeta_{0:\infty}^{*},\eta_{0:\infty}^{*})=V(x_0,i)<\infty\\
& \small{\implies} \mathbb{E}\Big[ \sum_{k=0}^{\infty} ||x_k||_{W(r_k)}^2+||u_k^*||_{\xi_{k}R(r_k)\xi_k}^2-\gamma^2 ||w_k||^2 \Big| \mathcal{I}_0 \Big] < \infty.
\end{split}
\end{equation*}
\normalsize
As $w_k\in \mathcal{L}_2([0,\infty),\mathbb{R}^s)$ or $ \sum_{k=0}^{\infty}||w_k||^2<\infty$,
\small
\begin{equation}
\begin{split}
& \hspace{0.7cm} \mathbb{E}\Big[ \sum_{k=0}^{\infty} ||x_k||_{W(r_k)}^2+||u_k^*||_{\mathbf{L}_k\big(\mathscr{Q}(r_k,.)\big)}^2 \Big| \mathcal{I}_0  \Big] < \infty \\
& \small{\implies}  \mathbb{E}\Big[ \sum_{k=0}^{\infty} \Big( ||x_k||_{W(r_k)}^2+||x_k||^2_{\tilde{\Gamma}^T_k \mathbf{L}_k\big(\mathscr{Q}(r_k,.)\big)\tilde{\Gamma}_k} \Big| \mathcal{I}_0  \Big] <\infty.
\end{split}
\label{stab_eq}
\end{equation}
\normalsize
Since $W(r_k)\geq 0$, $\mathbf{L}_k\big(\mathscr{Q}(r_k,.)\big)>0$ (as $R(r_k)>0$) for all $k$, $r_k\in \mathcal{D}$, (\ref{stab_eq}) guaranties the convergence of the infinite series $\sum_{k=0}^{\infty} \mathbb{E} \Big[ ||x_k||^2_{W(r_k)} \Big| \mathcal{I}_0  \Big]$ and  $\sum_{k=0}^{\infty} \mathbb{E} \Big[ ||x_k||^2_{\tilde{\Gamma}^T_k \mathbf{L}_k\big(\mathscr{Q}(r_k,.)\big)\tilde{\Gamma}_k} \Big| \mathcal{I}_0  \Big]$. In view of {Theorem 3.23} in \cite{rudin1976principles}, convergence of the infinite series  $\sum_{k=0}^{\infty} \mathbb{E} \Big[ ||x_k||^2_{\tilde{\Gamma}^T_k \mathbf{L}_k\big(\mathscr{Q}(r_k,.)\big)\tilde{\Gamma}_k} \Big| \mathcal{I}_0  \Big]$ implies that $\underset{k\rightarrow \infty}{\textrm{lim}}\mathbb{E} \Big[ ||x_k||^2_{\tilde{\Gamma}^T_k \mathbf{L}_k\big(\mathscr{Q}(r_k,.)\big)\tilde{\Gamma}_k} \Big| \mathcal{I}_0  \Big] = 0$. Hence, as  $\mathbf{L}_k\big(\mathscr{Q}(r_k,.)\big)>0$ for all $r_k \in \mathcal{D}$, one gets that $\underset{k\rightarrow \infty}{\textrm{lim}} \tilde{\Gamma}_{k} x_k = 0$ with probability 1. \\
We now claim that  $\underset{k\rightarrow \infty}{\textrm{lim}}\mathbb{E}\Big[ ||x_k||^2\Big| \mathcal{I}_0\Big]= 0$. We shall use contradiction in order to prove the claim. Assume that $\underset{k\rightarrow \infty}{\textrm{lim}}\mathbb{E}\Big[||x_k||^2\Big| \mathcal{I}_0 \Big]\neq 0$.\\
Using the system dynamics (\ref{system_eq}) with optimal control input $u_k^*=-\tilde{\Gamma}_kx_k$:
\small
\begin{equation}
\begin{split}
&  \underset{k\rightarrow \infty}{\textrm{lim}} \mathbb{E}\Big[ x_{k+1}^TW(r_{k+1})x_{k+1}\Big| \mathcal{I}_0
\Big]\\
& = \underset{k\rightarrow \infty}{\textrm{lim}} \Big{\{}  \mathbb{E}\Big[x_k^TA^T(r_k)W(r_{k+1})\Big(A(r_k)x_k-2B(r_k)\mathbf{L}_k\big(\mathcal{N}(.)\big)\tilde{\Gamma}_kx_k\\
&+2D_1(r_k)w_k\Big)-2x_k^T\tilde{\Gamma}^T_k\mathbf{L}_k\big(\mathcal{N}(.)\big)B^T(r_k)W(r_{k+1})D_1(r_k)w_k\\
& +x_k^T\tilde{\Gamma}^T_k\mathbf{L}_k\big(\mathcal{N}(.)\big)B^T(r_k)W(r_{k+1})B(r_k)\tilde{\Gamma}_kx_k\\
&+w_k^TD_1^T(r_k)W(r_{k+1})D_1(r_k)w_k \Big| \mathcal{I}_0 \Big] \Big{\}} .
\end{split}
\label{stab_1_1}
\end{equation}
\normalsize
As $\underset{p\rightarrow \infty}{\textrm{lim}} \tilde{\Gamma}_{k} x_k = 0$ with probability $1$ and $w_k \in \mathcal{L}_2([0,\infty),\mathbb{R}^s)$, from (\ref{system_eq}), $\underset{k\rightarrow \infty}{\textrm{lim}}\Big[x_{k+1}\Big]=\underset{k\rightarrow \infty}{\textrm{lim}}\Big[A(r_k)x_k\Big]$. Thus, (\ref{stab_1_1}) implies:
\small
\begin{equation}
\begin{split}
&  \underset{k\rightarrow \infty}{\textrm{lim}} \mathbb{E}\Big[ x_{k+1}^TW(r_{k+1})x_{k+1} \Big| \mathcal{I}_0 \Big]\\
& = \underset{k\rightarrow \infty}{\textrm{lim}} \mathbb{E}\Big[x_k^TA^T(r_k)W(r_{k+1})A(r_k)x_k\Big| \mathcal{I}_0\Big].
\end{split}
\label{stab_1_1_a}
\end{equation}
\normalsize
In similar fashion, we get:
\small
\begin{equation}
\begin{split}
&  \underset{k\rightarrow \infty}{\textrm{lim}} \mathbb{E}\Big[ x_{k+2}^TW(r_{k+2})x_{k+2}\Big| \mathcal{I}_0\Big] \\
& = \underset{k\rightarrow \infty}{\textrm{lim}} \mathbb{E}\Big[x_{k+1}^TA^T(r_{k+1})W(r_{k+2})A(r_{k+1})x_{k+1}\Big| \mathcal{I}_0\Big]\\
&= \underset{k\rightarrow \infty}{\textrm{lim}} \mathbb{E}\Big[x_{k}^TA^T(r_{k})A^T(r_{k+1})W(r_{k+2})A(r_{k+1})A(r_k)x_{k}\Big| \mathcal{I}_0\Big].
\end{split}
\label{stab_1_2}
\end{equation}
\normalsize
and so on.\\
% As the limit is taken as $p\rightarrow \infty$, one can manipulate the time-index $p$ such that the particular transition $\{r_p,r_{p+1},...,r_{\mathscr{F}}\}$ occurs almost surely such that $\textrm{rank} \hspace{0.1cm} \mathcal{O}(r_p,r_\mathscr{F})=n$.
%Now, if system (\ref{system_eq}) with $u_k\equiv 0$, $w_k\equiv 0$ is weakly observable then, using the similar line of argument as used in the proof for Lemma \ref{lem_q_multi}, 
From equations (\ref{stab_1_1_a}), (\ref{stab_1_2}),... one gets for $\mathscr{F}\geq n$:
\small
\begin{equation}
\begin{split}
& \underset{k\rightarrow \infty}{\textrm{lim}} \mathbb{E}\Big[ x_k^TW(r_k)x_k +x_{k+1}^TW(r_{k+1})x_{k+1}+ x_{k+2}^TW(r_{k+2})x_{k+2}\\
& + .............+ x_{k+\mathscr{F}}^TW(r_{k+\mathscr{F}})x_{k+\mathscr{F}}\Big| \mathcal{I}_0 \Big] \\
& = \underset{k\rightarrow \infty}{\textrm{lim}} \mathbb{E}\Big[ x_k^TW(r_k)x_k+x_k^TA^T(r_k)W(r_{k+1})A(r_k)x_k\\
& +x_k^TA^T(r_k)A^T(r_{k+1})W(r_{k+2})A(r_{k+1})A(r_k)x_k+...\\
&+ x_k^T\Big(\Pi_{l=k}^{{k+\mathscr{F}}}A(r_l)\Big)^TW(r_{k+\mathscr{F}})\Pi_{l=k}^{{k+\mathscr{F}}}A(r_l)x_k\Big| \mathcal{I}_0]\\
& = \underset{k\rightarrow \infty}{\textrm{lim}} \mathbb{E}\Big{[} x_k^T \mathscr{Y} \Big( r_k,r_{k+1},...,r_{k+\mathscr{F}}\Big) x_k\Big| \mathcal{I}_0
\Big{]}.
\end{split}
\label{stab_final}
\end{equation}
\normalsize
Note that the Markov chain $\{r_k\}$ is irreducible and has a finite numeber of states. Thus from Proposition \ref{prop_1}, it is recurrent. Hence, in light of Proposition \ref{prop_2}, it has a nonzero limiting distribution for all the states.  Therefore, one can prove the following inequality by using the similar line of argument as used in the proof for Lemma \ref{lem_q_multi}.
\small
\begin{equation}
\begin{split}
&\underset{k\rightarrow \infty}{\textrm{lim}} \mathbb{E}\Big{[} x_k^T \mathscr{Y} \Big( r_k,r_{k+1},...,r_{k+\mathscr{F}}\Big) x_k \Big| \mathcal{I}_0.
\Big{]}>0.
\label{stab_rt}
\end{split}
\end{equation}
\normalsize
Then, from (\ref{stab_final}) and (\ref{stab_rt}), one easily gets that, for certain $r_k$, $\underset{k\rightarrow \infty}{\textrm{lim}} \mathbb{E} \Big[ x_k^TW(r_k)x_k\Big| \mathcal{I}_0 \Big] \neq 0$. Therefore, considering {Theorem 3.23} in \cite{rudin1976principles}, we can infer that the infinite series $\sum_{k=0}^{\infty}  \mathbb{E} \Big[ x_k^TW(r_k)x_k\Big| \mathcal{I}_0\Big]$ does not converge.   
Further, $\mathbf{L}_k\Big(\mathscr{Q}(i,.)\Big)>0$ for all $k\in [0,\infty)$, $i\in \mathcal{D}$. Thus,
\small
\begin{equation*}
\mathbb{E}\Big[\sum_{k=0}^{\infty} ||x_k||^2_{W(r_k)}+ ||x_k||^2_{\tilde{\Gamma}^T_{k} \mathbf{L}_{k}\big(\mathscr{Q}(r_{k},.)\big)\tilde{\Gamma}_{k}}\Big| \mathcal{I}_0 \Big]\rightarrow \infty.
\end{equation*} 
\normalsize
Hence, we arrive at a contradiction. Therefore, $\underset{k\rightarrow \infty}{\textrm{lim}}\mathbb{E}\Big[||x_k||^2\Big| \mathcal{I}_0\Big]=0 $. \QEDB
\vspace{-0.5cm} 
  \section{Numerical Example}  
  Let us consider an MJLS with the following system parameters:
  \small
  \begin{equation*}
  \mathcal{D}=\{1,2\},
  \end{equation*}
 \begin{equation*}
 A(1)= \begin{aligned}
         \begin{bmatrix}
         1 \ $~~~~$ 2 \ $~~~~$ 1 \\
         0 \ $~~~~$  1 \ $~~~~$ 1\\
         1 \ $~~~~$ 0 \ $~~~~$ 2  
         \end{bmatrix}, 
         \end{aligned}
      \hspace{0.2cm}  B(1)= \begin{aligned}
          \begin{bmatrix}
         1 \ $~~~~$ 2 \\
         1 \ $~~~~$ 0 \\ 
         0 \ $~~~~$ 1   
         \end{bmatrix}
\end{aligned}, 
 \hspace{0.2cm}  D_1(1)= \begin{aligned}
          \begin{bmatrix}
         1  \\
         1  \\ 
         1   
         \end{bmatrix},
\end{aligned}
\end{equation*}
\begin{equation*}
 C(1)= \begin{aligned}
         \begin{bmatrix}
         0 \ $~~~~$ 0 \ $~~~~$ 0 \\
         0 \ $~~~~$  0 \ $~~~~$ 0\\
         1 \ $~~~~$ 1 \ $~~~~$ 1  
         \end{bmatrix}, 
         \end{aligned}
      \hspace{0.2cm}  D(1)= \begin{aligned}
          \begin{bmatrix}
         1 \ $~~~~$ 0 \\
         0 \ $~~~~$ 1 \\ 
         0 \ $~~~~$ 0   
         \end{bmatrix}
\end{aligned}, 
\end{equation*}
\begin{equation*} 
 A(2)= \begin{aligned}
         \begin{bmatrix}
         1 \ $~~~~$ 0 \ $~~~~$ 1 \\
         0 \ $~~~~$  1 \ $~~~~$ 0\\
         1 \ $~~~~$ 0 \ $~~~~$ 2  
         \end{bmatrix}, 
         \end{aligned}
      \hspace{0.2cm}  B(2)= \begin{aligned}
          \begin{bmatrix}
         1 \ $~~~~$ 2 \\
         1 \ $~~~~$ 0 \\ 
         0 \ $~~~~$ 1   
         \end{bmatrix}
\end{aligned}, 
 \hspace{0.2cm}  D_1(2)= \begin{aligned}
          \begin{bmatrix}
         1  \\
         1  \\ 
         1   
         \end{bmatrix},
\end{aligned}
\end{equation*}
\begin{equation*}
 C(2)= \begin{aligned}
         \begin{bmatrix}
         0 \ $~~~~$ 0 \ $~~~~$ 0 \\
         0 \ $~~~~$  0 \ $~~~~$ 0\\
         1 \ $~~~~$ 1 \ $~~~~$ 1  
         \end{bmatrix}, 
         \end{aligned}
      \hspace{0.2cm}  D(2)= \begin{aligned}
          \begin{bmatrix}
         1 \ $~~~~$ 1 \\
         0 \ $~~~~$ 1 \\ 
         0 \ $~~~~$ 0   
         \end{bmatrix},
\end{aligned}
\end{equation*} 
\normalsize
State transition matrix for the Markov chain $\{r_k\}$ is given by:
\begin{equation*}
\mathcal{T}=
\begin{aligned}
\begin{bmatrix}
0.45 \ $~~~~$ 0.55 \\
0.4 \ $~~~~~$  0.6  
\end{bmatrix}. 
\end{aligned}
\end{equation*}
\normalsize
Also, consider the following parameters:
\small
\begin{equation*}
W(1)= W(2) = \begin{aligned}
         \begin{bmatrix}
         1 \ $~~~~$ 1 \ $~~~~$ 1 \\
         1 \ $~~~~$  1 \ $~~~~$ 1\\
         1 \ $~~~~$ 1 \ $~~~~$ 1  
         \end{bmatrix}, 
         \end{aligned}
\end{equation*}
and 
\begin{equation*}
R(1)=  \begin{aligned}
          \begin{bmatrix}
         1 \ $~~~~$ 0 \\
         0 \ $~~~~$ 1 
         \end{bmatrix}
\end{aligned}, \hspace{0.2cm} 
R(2)=  \begin{aligned}
          \begin{bmatrix}
         1 \ $~~~~$ 1 \\
         1 \ $~~~~$ 2 
         \end{bmatrix}.
\end{aligned}
\end{equation*} 
\normalsize
%\begin{figure} [!htp]
%{\centering
%\includegraphics[scale=0.26]{convergence_single_march} 
%\caption{Behaviour of the optinal cost $J_N(\zeta^{*N-1},\eta^{*N-1})$ with $\alpha =0.9$, $\beta =0.1$ (blue graph) and $\alpha =0.89$, $\beta =0.1$ (red graph). } 
%\label{fig:converge_signle}
%} 
%\end{figure}
\begin{comment}
\begin{figure} [!htp]
{\centering
\includegraphics[scale=0.26]{divergence_single_march} 
\caption{Behaviour of the optinal cost $J_N(\zeta^{*N-1},\eta^{*N-1})$ with $\alpha =0.85$ and $\beta =0.1$. }
\label{fig:diverge_signle} 
} 
\end{figure}
\begin{figure} [!htp]
{\centering
\includegraphics[scale=0.36]{alpha_Vs_gamma_c_single_channel} 
\caption{ Behaviour of $\gamma_c$ for different $\alpha$, and $\beta =0.1$. } 
\label{fig:alpha_vs_gamma_signle}
} 
\end{figure}
\begin{figure} [!htp]
{\centering
\includegraphics[scale=0.36]{beta_Vs_gamma_c_single_channel} 
\caption{ Behaviour of $\gamma_c$ for different $\beta$, and $\alpha =0.89$. } 
\label{fig:beta_vs_gamma_signle}
} 
\end{figure}
\begin{figure} [!htp]
{\centering
\includegraphics[scale=0.26]{stab_single_march} 
\caption{Response of the states with the optimal control law for $\alpha = 0.9$ and $\beta = 0.11$. } 
\label{fig:stab_signle}
} 
\end{figure}
\end{comment}
\begin{figure} [!htp]
{\centering
\includegraphics[scale=0.26]{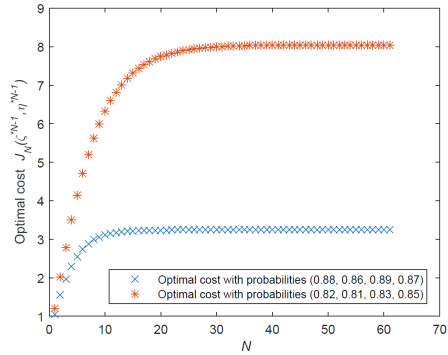} 
\caption{Optimal cost $J_N(\zeta^{*N-1},\eta^{*N-1})$ with $\bar{v}^1=0.88$, $\bar{v}^2=0.86$, $\bar{\mu}^1=0.89$, $\bar{\mu}^2=0.87$ (blue graph) and  $\bar{v}^1=0.82$, $\bar{v}^2=0.81$, $\bar{\mu}^1=0.83$, $\bar{\mu}^2=0.85$ (red graph).} 
\label{fig:converge_multi}
} 
\end{figure}
\begin{figure} [!htp]
{\centering
\includegraphics[scale=0.26]{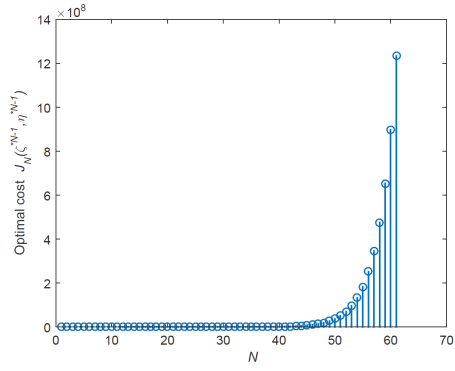} 
\caption{Optimal cost $J_N(\zeta^{*N-1},\eta^{*N-1})$ with $\bar{v}^1=0.72$, $\bar{v}^2=0.76$, $\bar{\mu}^1=0.77$, $\bar{\mu}^2=0.67$.} 
\label{fig:diverge_multi}
} 
\end{figure} 

\begin{figure} [!htp]
{\centering
\includegraphics[scale=0.35]{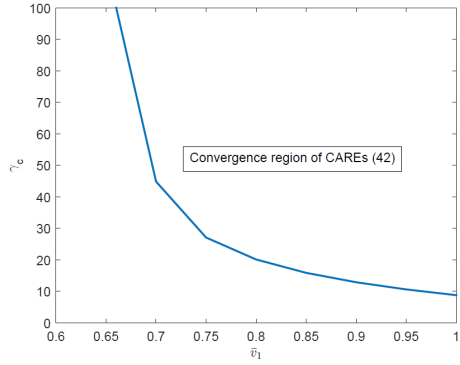} 
\caption{Variation of $\gamma_c$ with different $\bar{v}^1$, and $\bar{v}^2=0.85$, $\bar{\mu}^1=0.82$, $\bar{\mu}^2=0.8$.} 
\label{fig:v1_vs_gamma_signle}
}
\end{figure} 
\begin{figure} [!htp]
{\centering
\includegraphics[scale=0.36]{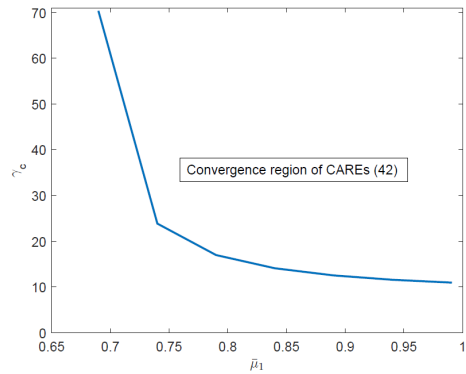} 
\caption{Variation of $\gamma_c$ with different $\bar{\mu}^1$, and $\bar{v}^1=0.85$, $\bar{v}^2=0.83$, $\bar{\mu}^2=0.82$.} 
\label{fig:n1_vs_gamma_signle}
} 
\end{figure}
\begin{figure} [!htp]
{\centering
\includegraphics[scale=0.27]{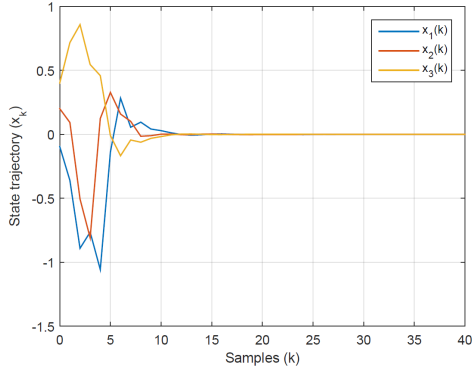} 
\caption{Response of the states with the optimal control law for $\bar{v}^1=0.81$, $\bar{v}^2=0.8$, $\bar{\mu}^1=0.81$, $\bar{\mu}^2=0.79$. } 
\label{fig:stab_multi}
} 
\end{figure}
\vspace{-0.5cm}
\par In order to demonstrate the influence of the control packet arrival probabilities on the infinite horizon optimal cost, we compute the optimal cost for different horizon. With $x_0=[0.1 \hspace{0.15cm} 0.2 \hspace{0.15cm} 0.3]^T$ as the initial state vector, optimal cost is computed using (\ref{eq_V_0}). Fig.~\ref{fig:converge_multi} shows the behavior of the optimal cost $J_N(\zeta^{*N-1}, \eta^{*N-1})$ with the probabilities  $\bar{v}^1=0.88$, $\bar{v}^2=0.86$, $\bar{\mu}^1=0.89$, $\bar{\mu}^2=0.87$ and $\bar{v}^1=0.82$, $\bar{v}^2=0.81$, $\bar{\mu}^1=0.83$, $\bar{\mu}^2=0.85$. One can observe that as packet arrival probabilities are reduced, the optimal cost $J_N(\zeta^{*N-1}, \eta^{*N-1})$ converges to a higher value. If the probabilities are  reduced to $\bar{v}^1=0.72$, $\bar{v}^2=0.76$, $\bar{\mu}^1=0.77$, $\bar{\mu}^2=0.67$, it can be observed, as shown in Fig.~\ref{fig:diverge_multi}, that the optimal cost $J_N(\zeta^{*N-1}, \eta^{*N-1})$ does not converge. In Fig.~\ref{fig:v1_vs_gamma_signle}, the dependence of the critical value of the $H_{\infty}$ disturbance attenuation level $\gamma_c$ on $\bar{v}^1$ is demonstrated while keeping $\bar{v}^2=0.85$, $\bar{\mu}^1=0.82$ and $\bar{\mu}^2=0.8$. Similarly, Fig.~\ref{fig:n1_vs_gamma_signle} shows the variation of $\gamma_c$ with respect to $\bar{\mu}^1$ while keeping $\bar{v}^1=0.85$, $\bar{v}^2=0.83$, $\bar{\mu}^2=0.82$. The regions above the curve in Fig.~\ref{fig:v1_vs_gamma_signle} and Fig.~\ref{fig:n1_vs_gamma_signle} correspond to the feasible region for convergence of the CAREs (\ref{care_multi}).  Fig.~\ref{fig:v1_vs_gamma_signle} and   Fig.~\ref{fig:n1_vs_gamma_signle} suggest that as the control packet arrival probabilities ($\bar{v}^1$ and $\bar{\mu}^1$, respectively) reduce, $\gamma_c$ goes on increasing. Further, below a critical value of the control packet arrival probability, there does not exists any finite value for $\gamma_c$ such that CAREs (\ref{care_multi_infi}) admit a unique fixed-point solution. The decaying state responses with the optimal controller is shown in Fig. ~\ref{fig:stab_multi} with disturbance input $w_k=sin(0.2\pi k)cos(0.2\pi k)e^{-k/2}$. 
  %Let us consider a scalar system given by following system parameters: $\mathcal{D}=\{1,2\}$, $A(1)=2$, $A(2)=2.1$, $B(1)=B(2)=1$, $D_1(1)=D_2(2)=1$, $W(1)=W(2)=1$, $R(1)=R(2)=1$. 
  %\begin{figure} [h!]
%{\centering
%\includegraphics[scale=0.5]{Gamma_c_Vs_alpha} 
%\caption{Variation of critical $\gamma_c$ w.r.t.  $\alpha$ for $\beta=0.15$ } 
%} 
%\end{figure}
 % \begin{figure} [h!]
%{\centering
%\includegraphics[scale=0.5]{Gamma_c_Vs_beta} 
%\caption{Variation of critical $\gamma_c$ w.r.t.  $\beta$ for $\alpha=0.82$ } 
%} 
%\end{figure}
\vspace{-0.1cm}
\section{Conclusions}
In this paper, we have designed the optimal $H_{\infty}$ controller for a Markovian jumped linear system over multiple communication channels. Existence conditions for the finite-horizon controller are derived. It is observed that the convergence of the infinite-horizon cost function depends on the control packet arrival probabilities. Stability of the closed-loop system with the optimal controller in the face of random packet loss has also been established.  
\vspace{-0.2cm}
\bibliographystyle{IEEEtran}
\bibliography{abhijit_ref} 

\begin{thebibliography}{10}
\providecommand{\url}[1]{#1}
\csname url@samestyle\endcsname
\providecommand{\newblock}{\relax}
\providecommand{\bibinfo}[2]{#2}
\providecommand{\BIBentrySTDinterwordspacing}{\spaceskip=0pt\relax}
\providecommand{\BIBentryALTinterwordstretchfactor}{4}
\providecommand{\BIBentryALTinterwordspacing}{\spaceskip=\fontdimen2\font plus
\BIBentryALTinterwordstretchfactor\fontdimen3\font minus
  \fontdimen4\font\relax}
\providecommand{\BIBforeignlanguage}[2]{{%
\expandafter\ifx\csname l@#1\endcsname\relax
\typeout{** WARNING: IEEEtran.bst: No hyphenation pattern has been}%
\typeout{** loaded for the language `#1'. Using the pattern for}%
\typeout{** the default language instead.}%
\else
\language=\csname l@#1\endcsname
\fi
#2}}
\providecommand{\BIBdecl}{\relax}
\BIBdecl

\bibitem{oncu2012string}
S.~Oncu, N.~Van~de Wouw, W.~M.~H. Heemels, and H.~Nijmeijer, ``String
  {S}tability of {I}nterconnected {V}ehicles {U}nder {C}ommunication
  {C}onstraints,'' in \emph{Proc. 51st IEEE CDC}, 2012, pp. 2459--2464.

\bibitem{millan2014formation}
P.~Mill{\'a}n, L.~Orihuela, I.~Jurado, and F.~R. Rubio, ``Formation {C}ontrol
  of {A}utonomous {U}nderwater {V}ehicles {S}ubject to {C}ommunication
  {D}elays,'' \emph{IEEE Trans. Contr. Syst. Tech.}, vol.~22, no.~2, pp.
  770--777, 2014.

\bibitem{yao2015wide}
W.~Yao, L.~Jiang, J.~Wen, Q.~Wu, and S.~Cheng, ``Wide-{A}rea {D}amping
  {C}ontroller for {P}ower {S}ystem {I}nterarea {O}scillations: A {N}etworked
  {P}redictive {C}ontrol {A}pproach,'' \emph{IEEE Trans. Contr. Syst. Tech.},
  vol.~23, no.~1, pp. 27--36, 2015.

\bibitem{f}
W.~Zhang, M.~S. Branicky, and S.~M. Phillips, ``Stability of {N}etworked
  {C}ontrol {S}ystems,'' \emph{IEEE Contr. Syst. Mag.}, vol.~21, no.~1, pp.
  84--99, 2001.

\bibitem{g}
L.~Schenato, B.~Sinopoli, M.~Franceschetti, K.~Poolla, and S.~S. Sastry,
  ``Foundations of control and estimation over lossy networks,'' \emph{Proc.
  IEEE}, vol.~95, no.~1, pp. 163--187, 2007.

\bibitem{h}
W.-W. Che, J.-L. Wang, and G.-H. Yang, ``Quantised ${H}_{\infty}$ filtering for
  networked systems with random sensor packet losses,'' \emph{IET Contr. Theory
  App.}, vol.~4, no.~8, pp. 1339--1352, 2010.

\bibitem{22}
J.~Moon and T.~Ba{\c{s}}ar, ``Control over {TCP}-like lossy networks: {A}
  dynamic game approach,'' in \emph{Proc. Amer. Contr. Conf.}, 2013, pp.
  1578--1583.

\bibitem{j}
Y.~Mo, E.~Garone, and B.~Sinopoli, ``{LQG} control with {M}arkovian packet
  loss,'' in \emph{Proc. Euro. Contr. Conf. (ECC)}, 2013, pp. 2380--2385.

\bibitem{vargas2016optimal}
A.~N. Vargas, L.~P. Sampaio, L.~Acho, L.~Zhang, and J.~B. do~Val, ``Optimal
  control of dc-dc buck converter via linear systems with inaccessible
  markovian jumping modes,'' \emph{IEEE Trans. Control Syst. Tech.}, vol.~24,
  no.~5, pp. 1820--1827, 2016.

\bibitem{costa2006discrete}
O.~L.~V. Costa, M.~D. Fragoso, and R.~P. Marques, \emph{Discrete-time {M}arkov
  jump linear systems}.\hskip 1em plus 0.5em minus 0.4em\relax Springer Science
  \& Business Media, 2006.

\bibitem{bruno}
B.~Sinopoli, L.~Schenato, M.~Franceschetti, K.~Poolla, and S.~S. Sastry,
  ``Optimal control with unreliable communication: the {TCP} case,'' in
  \emph{Proc. Amer. Contr. Conf.}, 2005, pp. 3354--3359.

\bibitem{garone}
E.~Garone, B.~Sinopoli, A.~Goldsmith, and A.~Casavola, ``{LQG} control for
  {MIMO} systems over multiple erasure channels with perfect acknowledgment,''
  \emph{IEEE Trans. Autom. Contr.}, vol.~57, no.~2, pp. 450--456, 2012.

\bibitem{abhijit_mtns}
A.~Mazumdar, S.~Krishnaswamy, and S.~Majhi, ``Linear {Q}uadratic {O}ptimal
  {C}ontrol of {J}ump {S}ystem over {M}ultiple {E}rasure {C}hannels,'' in
  \emph{23rd Inter. {S}ymp. {M}ath. {T}heory {N}etw. {S}yst. ({MTNS}), Hong
  Kong}, 2018.

\bibitem{p}
P.~Seiler and R.~Sengupta, ``An ${H}_{\infty}$ approach to networked control,''
  \emph{IEEE Trans. Auto. Contr.}, vol.~50, no.~3, pp. 356--364, 2005.

\bibitem{wang2007robust}
Z.~Wang, F.~Yang, D.~W. Ho, and X.~Liu, ``Robust ${H}_{\infty}$ {C}ontrol for
  {N}etworked {S}ystems {W}ith {R}andom {P}acket {L}osses,'' \emph{IEEE Trans.
  Syst. Man Cyber. Part B (Cybernetics)}, vol.~37, no.~4, pp. 916--924, 2007.

\bibitem{18}
H.~Ishii, ``${H}_{\infty}$ control with limited communication and message
  losses,'' \emph{Syst. Contr. Let.}, vol.~57, no.~4, pp. 322--331, 2008.

\bibitem{wang2013h}
D.~Wang, J.~Wang, and W.~Wang, ``${H}_{\infty}$ controller design of networked
  control systems with {M}arkov packet dropouts,'' \emph{IEEE trans. syst. man
  cyber.: Syst.}, vol.~43, no.~3, pp. 689--697, 2013.

\bibitem{i}
J.~Moon and T.~Ba{\c{s}}ar, ``Control over lossy networks: {A} dynamic game
  approach,'' in \emph{Proc. Amer. Contr. Conf. (ACC)}, 2014, pp. 5367--5372.

\bibitem{m}
J.~Moon and T.~Basar, ``Minimax control over unreliable communication
  channels,'' \emph{Automatica}, vol.~59, pp. 182--193, 2015.

\bibitem{moon2014minimax}
J.~Moon and T.~Ba{\c{s}}ar, ``Minimax control of {MIMO} systems over multiple
  {TCP}-like lossy networks,'' \emph{IFAC Proceedings Volumes}, vol.~47, no.~3,
  pp. 110--115, 2014.

\bibitem{mazumdar2017h}
A.~Mazumdar, S.~Krishnaswamy, and S.~Majhi, ``${H}_{\infty}$-optimal {C}ontrol
  over erasure channel,'' \emph{IFAC-PapersOnLine}, vol.~50, no.~1, pp.
  349--354, 2017.

\bibitem{liu2018dynamic}
X.~Liu, G.~Ma, P.~R. Pagilla, and S.~S. Ge, ``Dynamic output feedback
  asynchronous control of networked {M}arkovian jump systems,'' \emph{IEEE
  Trans. Syst. Man Cybern.: Syst.}, 2018.

\bibitem{kawan2016network}
C.~Kawan and J.-C. Delvenne, ``Network entropy and data rates required for
  networked control,'' \emph{IEEE Trans. Contr. Netw. Syst.}, vol.~3, no.~1,
  pp. 57--66, 2016.

\bibitem{l}
G.~Ha{\ss}linger and O.~Hohlfeld, ``The {G}ilbert-{E}lliott model for packet
  loss in real time services on the internet,'' in \emph{Meas. Model. Eval.
  Comp. Comm. Syst. (MMB), 2008 14th GI/ITG Conf.}\hskip 1em plus 0.5em minus
  0.4em\relax VDE, 2008, pp. 1--15.

\bibitem{ji1988controllability}
Y.~Ji and H.~J. Chizeck, ``Controllability, observability and discrete-time
  {M}arkovian jump linear quadratic control,'' \emph{Inter. Jour. Contr.},
  vol.~48, no.~2, pp. 481--498, 1988.

\bibitem{n}
T.~Ba{\c{s}}ar and P.~Bernhard, \emph{$H_{\infty}$ optimal control and related
  minimax design problems: a dynamic game approach}.\hskip 1em plus 0.5em minus
  0.4em\relax Springer Science \& Business Media, 2008.

\bibitem{rudin1976principles}
W.~Rudin, \emph{Principles of mathematical analysis}.\hskip 1em plus 0.5em
  minus 0.4em\relax McGraw-hill New York, 1976, vol.~3, no. 4.2.

\bibitem{iosifescu2014finite}
M.~Iosifescu, \emph{Finite Markov processes and their applications}.\hskip 1em
  plus 0.5em minus 0.4em\relax Courier Corporation, 2014.

\bibitem{ross2014introduction}
S.~M. Ross, \emph{Introduction to probability models}.\hskip 1em plus 0.5em
  minus 0.4em\relax Academic press, 2014.

\end{thebibliography}

\end{document}